\documentclass[journal, twocolumn, 10pt]{IEEEtran}
\usepackage{amsmath,amsfonts,amssymb,bm,graphicx, mathtools}
\usepackage[usenames,dvipsnames]{color}
\usepackage[noadjust]{cite}
\usepackage{verbatim, enumerate, paralist}%
\usepackage{epsfig, epstopdf}
\usepackage{algorithmicx, algpseudocode}
\usepackage{float, setspace}
\usepackage{url, caption, subcaption}
\usepackage{tikz, pgfkeys, pgfplots}
\usepackage[normalem]{ulem}

\usetikzlibrary{arrows,shapes,calc,positioning,plotmarks}

\DeclarePairedDelimiter{\ceil}{\lceil}{\rceil}
\DeclareMathOperator{\Tr}{Tr}

\makeatletter
\floatstyle{ruled}
\newfloat{algorithm}{tbp}{loa}
\providecommand{\algorithmname}{Algorithm}
\floatname{algorithm}{\protect\algorithmname}

\makeatother

\long\def\comment#1{}

\newcommand{\beq}{\begin{equation*}}
\newcommand{\eeq}{\end{equation*}}

\newfont{\bbb}{msbm10 scaled 700}

\newfont{\bb}{msbm10 scaled 1100}

\newcommand{\av}{{\bf a}}

\newcommand{\hv}{{\bf h}}

\newcommand{\xv}{{\bf x}}
\newcommand{\yv}{{\bf y}}

\newcommand{\zerov}{{\bf 0}}

\newcommand{\Am}{{\bf A}}

\newcommand{\Cm}{{\bf C}}

\newcommand{\Fm}{{\bf F}}

\newcommand{\Id}{{\bf I}}

\newcommand{\Pm}{{\bf P}}

\newcommand{\Rm}{{\bf R}}

\newcommand{\Xm}{{\bf X}}

\newcommand{\Ac}{{\cal A}}

\newcommand{\Cc}{{\cal C}}

\newcommand{\Hc}{{\cal H}}

\newcommand{\Nc}{{\cal N}}

\newcommand{\Pc}{{\cal P}}

\newcommand{\Rc}{{\cal R}}
\newcommand{\Sc}{{\cal S}}

\newcommand{\Wc}{{\cal W}}

\newcommand{\Xc}{{\cal X}}
\newcommand{\Yc}{{\cal Y}}
\newcommand{\Zc}{{\cal Z}}

\newcommand{\Hbc}{{\bm{\Hc}}}

\newcommand{\Wbc}{{\bm{\Wc}}}

\newcommand{\Ybc}{{\bm{\Yc}}}
\newcommand{\Zbc}{{\bm{\Zc}}}

\newcommand{\herm}{{\sf H}}

\newcommand{\transp}{{\sf T}}

\title{Efficient Coordinated Recovery of Sparse Channels in Massive MIMO}
\author{Mudassir~Masood, Laila H. Afify, and~Tareq Y. Al-Naf{}fouri*,~\IEEEmembership{Member,~IEEE,}
\thanks{Authors are with the Department
of Electrical Engineering, King Abdullah University of Science \& Technology, Thuwal 23955-6900, Kingdom of Saudi Arabia,
e-mail: mudassir.masood@kaust.edu.sa, laila.afify@kaust.edu.sa, tareq.alnaffouri@kaust.edu.sa}
\thanks{Tareq Y. Al-Naf{}fouri is also associated with the Department of Electrical Engineering, King Fahd University of Petroleum and Minerals, Dhahran 31261, Kingdom of Saudi Arabia.}
}

\begin{document}
\maketitle

\begin{abstract}
This paper addresses the problem of estimating sparse channels in massive MIMO-OFDM systems. Most wireless channels are sparse in nature with large delay spread. In addition, these channels as observed by multiple antennas in a neighborhood have  approximately common support. The sparsity and common support properties are attractive when it comes to the efficient estimation of large number of channels in massive MIMO systems. Moreover, to avoid pilot contamination and to achieve better spectral efficiency, it is important to use a small number of pilots. We present a novel channel estimation approach which utilizes the sparsity and common support properties to estimate sparse channels and require a small number of pilots. Two algorithms based on this approach have been developed which perform Bayesian estimates of sparse channels even when the prior is non-Gaussian or unknown. Neighboring antennas share among each other their beliefs about the locations of active channel taps to perform estimation. The coordinated approach improves channel estimates and also reduces the required number of pilots. Further improvement is achieved by the data-aided version of the algorithm. Extensive simulation results are provided to demonstrate the performance of the proposed algorithms.
\end{abstract}
\section{Introduction}
\label{sec:intro}

The deployment of multiple antennas in a wireless communication system offers key advantages to its performance in terms of power gain, channel robustness, diversity, and spatial multiplexing. The use of multiple antennas in rich scattering environments provides effective utilization of the scarcely available spectrum resources. As a result, multiple-input-multiple-output (MIMO) technology has gained much interest in the research community. 

Installing extra antennas to a MIMO system can introduce substantial enhancements in both link reliability and data throughput of the system \cite{LTEAdvanced}. Specifically, the use of very large antenna arrays (in the order of multiples of a hundred antennas) 
has been found to be beneficial to overcome problems encountered in traditional MIMO settings. Such systems, known as \emph{massive} MIMO or \emph{large-scale} MIMO \cite{scaling_up},\cite{how_many_antennas}, also have the potential to scale down the transmission power because of the use of small active antennas with very low power. Moreover, in massive MIMO systems fast fading is averaged out and intracell interference almost vanishes. Thus, using large antenna arrays can play a key role in exploiting the true potential of traditional MIMO systems while at the same time overcoming several challenges. Massive MIMO is therefore considered as an emerging key technology that can meet the growing demands of current wireless systems. For interested readers, some other advantages of adding more antennas to the base station (BS) have been discussed in \cite{training_MU_MIMO, 6736761, non_cooperative_BS_antennas}.

In order to benefit from the advantages of massive MIMO systems, we need to determine the channel impulse response (CIR) for each transmit-receive link. In a typical massive MIMO system, a BS is equipped with a large antenna array and communicates with several users resulting in a large number of channels that need to be estimated. This results in a substantial increase in complexity which causes performance limitations at the BS. 
Obviously, obtaining CIR requires training data (pilots) to be sent by the users. It is known that the number of pilot symbols required is proportional to the total number of users \cite{training_MU_MIMO}. Therefore, as the number of users increases, there is a higher chance that the pilot sequences in the neighboring cells interfere with each other. This pilot contamination problem is a major limiting factor for the massive MIMO systems \cite{5898372, 5947131}. However, pilot contamination could be reduced if the reserved number of pilot tones is reduced. Therefore, in a multi-user scenario there is a need to reduce the number of pilots without affecting the CIR quality. Hence the development of efficient channel estimation techniques for massive MIMO that are computationally less complex and require less number of pilots is a challenge that needs to be thoroughly addressed.

Massive MIMO channel estimation is similar to the MIMO channel estimation. Existing literature includes several methods proposed for channel estimation in MIMO systems \cite{medles1, semi_blind_Swamy, semi_blind_MIMO_OFDM,6288608,5670986}. However, it is difficult to directly adopt these approaches for a number of reasons. For example,
\begin{enumerate}
\item There is a need to reduce the number of pilots.
\item All received (thousands of) signals in a massive MIMO system can not be processed efficiently at one central processor. Therefore, there is a need for methods/algorithms which are
    \begin{inparaenum}[\itshape a\upshape)]
    \item distributed;
    \item computationally efficient; and
    \item require little communication overhead.
    \end{inparaenum}
\item The antenna arrays could spread over a large space making it quite different in its model than a regular compact MIMO receiver.
\end{enumerate}

Recent works have indicated increased interest in the problem of massive MIMO channel estimation (see for example \cite{2014arXiv1401.5703S, 6415397, 5947131, 5898372, 6555020}). Most of these algorithms make use of the channel statistics. However, these statistics are usually not known and therefore some kind of assumption is made about the distribution of channel taps. Moreover, some of the techniques involve computationally expensive operations like inversion of channel covariance matrices which is not reasonable for the massive MIMO scenario.

It is well known that many wireless channels have impulse response that is sparse in the sense that they have very few significant paths. For example, see \cite{387095, barbotin2011estimating, semi_blind_MIMO_OFDM, 771349, wireless_sparse3, 4202180, 4042341, Rappaport2002wireless, 5670986, comp_channel_sensing} and the references therein. We would also like to add that in massive MIMO since a large number of antennas has to be placed usually it is difficult due to several space, structural, aesthetic constraints that antennas are positioned far from each other. Antenna separation is bound to decrease as we increase the number of antennas. This means that for antennas that are close to each other the times of arrival  will be similar however the amplitudes and phases of the paths will be different, implying common support. The ideas that we will put forward in this paper take advantage of the above-mentioned two properties.

In this paper, we propose a set of algorithms for channel estimation in massive MIMO. Specifically, we consider a base-station equipped with a large number of antennas serving several single-antenna user-equipments (UE). Our approach makes use of the fact that the wireless channels between a UE and base-station antennas are expected to be sparse and that neighboring antennas observe channels with similar support (i.e., sparsity pattern) but not necessarily the same fading along the active taps. The antennas share information with their neighbors to reach a decision on the most probable support. Decisions are made in a distributed manner with low complexity and communication overhead. In summary, the set of algorithms we propose in this paper has the following distinctive features:
\begin{enumerate}
\item It utilizes the sparsity of the CIR and the fact that channel supports for neighboring antennas are approximately the same.
\item It is Bayesian in nature. It utilizes the sparsity of CIR and acknowledges the Gaussianity of the additive noise but is agnostic to the distribution of the active taps of the CIR.
\item It has a distributed nature requiring limited communication between neighboring antennas. One version of the algorithm requires only integer communication between antennas.
\item It has a data-aided extension that identifies reliable carriers and uses them to further reduce the number of pilots and enhance the CIR estimate.
\end{enumerate}

The distributed Bayesian algorithm we develop in this paper is based on the Support Agnostic Bayesian Matching Pursuit algorithm (SABMP) developed by the authors in \cite{sabmp}.

The remainder of this paper is organized as follows. In Section \ref{sec:sysmodel}, we present the system model and formulate the channel estimation problem. In Section \ref{sec:sabmp}, we present a simple channel recovery method and propose enhancements to it that are required for the development of our coordinated recovery algorithms proposed in Section \ref{sec:coordinated}. A data-aided version of this algorithm is presented in Section \ref{sec:data-aided}. Simulation results are discussed in Section \ref{sec:results} and Section \ref{sec:conclusions} concludes the paper. 

\subsection{Notation}
We denote vectors with small-case bold-face letters (e.g., $\xv$), matrices with upper-case, bold-face letters (e.g., $\Xm$), and reserve calligraphic notation for symbols in frequency domain (e.g., $\bm{\Xc}$). We use $\xv_i$ to denote the $i^{th}$ column of matrix $\Xm$ and $x(j)$ to denote the $j^{th}$ entry of vector $\xv$. We also use $\Xm_{\Sc}$ to denote the sub-matrix formed by the columns $\{\xv_i : i \in \Sc \}$, indexed by the set $\Sc$. We use $\widehat{\xv}$, $\xv^*$, $\xv^\transp$, and $\xv^\herm$ to respectively denote the estimate, conjugate, transpose, and conjugate transpose of the vector $\xv$. Finally we use $\mathrm{diag}(\xv)$ to transform the vector $\xv$ into a diagonal matrix with the entries of $\xv$ spread along the diagonal.


\section{System Model and Problem Formulation}
\label{sec:sysmodel}
\subsection{Transmission Model}

We consider a MIMO-OFDM system in which the BS is equipped with a large two-dimensional antenna array consisting of $R=M \times G$ antennas distributed across $M$ rows and $G$ columns.\footnote{Depending on the value of $M$ and $G$, the antennas could have a linear or a rectangular configuration. Further, we would like to stress that while we confine our attention to rectangular configurations for convenience, our approach applies to any one-, two- or three-dimensional configuration of antennas as explained at the end of Sec. \ref{sec:data-aided}.} The base station serves a number of single-antenna terminals. Orthogonal frequency division multiplexing (OFDM) is adopted as the signaling mechanism. In an OFDM system, serially incoming bits are divided into $N$ parallel streams and mapped into a $Q$-ary QAM alphabet $\{ \Ac_1, \Ac_2, \dots, \Ac_{Q}\}$. This results in an $N$-dimensional data vector $\bm{\Xc} = \left[  \Xc(1), \Xc(2), \dots,\Xc(N)\right]^\transp$. The equivalent time-domain signal $\xv$  is obtained by taking the inverse Fourier transform of the data vector, i.e.,
\begin{align}
\xv&=\Fm^\herm \bm{\Xc},
\end{align}
where $\Fm$ is an $N \times N$ unitary discrete Fourier transform (DFT) matrix whose $(k,l)$ entry is given by
\begin{align}
f_{k,l}&=\frac{1}{\sqrt{N}} \exp{(-\jmath\frac{2\pi}{N}kl)}.
\end{align}
A cyclic prefix is inserted at the beginning of each symbol and the resulting signal is transmitted.


\subsection{Channel Model}\label{sec:channelmodel}
\begin{figure}[htbp]
 \centerline{ \includegraphics[width=0.9\columnwidth] {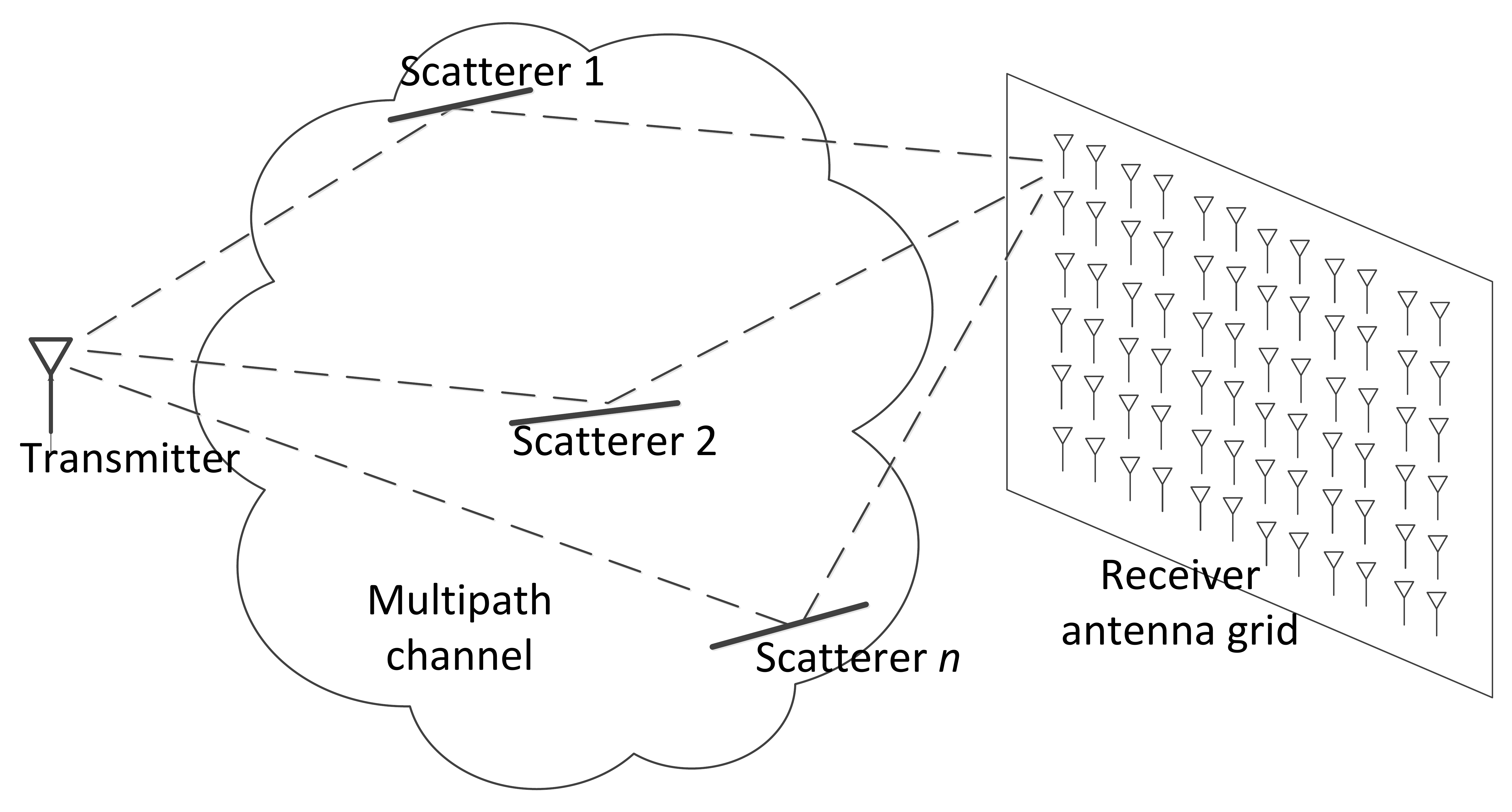}
 } \caption{In most wireless channels scatterers are sparsely distributed and the resulting channel impulse response is sparse.}
 \label{fig:scatterers}
 \end{figure}

It is known that most wireless channels can be modeled as discrete multipath channels with large delay spread and very few significant paths as scatterers are sparsely distributed in space (see Fig. \ref{fig:scatterers}). This makes the CIR sparse  \cite{wireless_sparse1,wireless_sparse2,Rappaport2002wireless}. Thus, for each transmit-receive link, we need only estimate a few significant multipath channel gains, which has the potential to reduce the pilot overhead substantially. We explicitly mention the sparsity property as property 1.

\begin{center}
\fbox{%
  \parbox{0.9\columnwidth}{%
    \begin{center}
        \textbf{Property 1:} \emph{The channel impulse response is sparse.}
    \end{center}
  }%
}
\end{center}

\begin{figure}[htbp]
\centering
\includegraphics[scale=0.75]{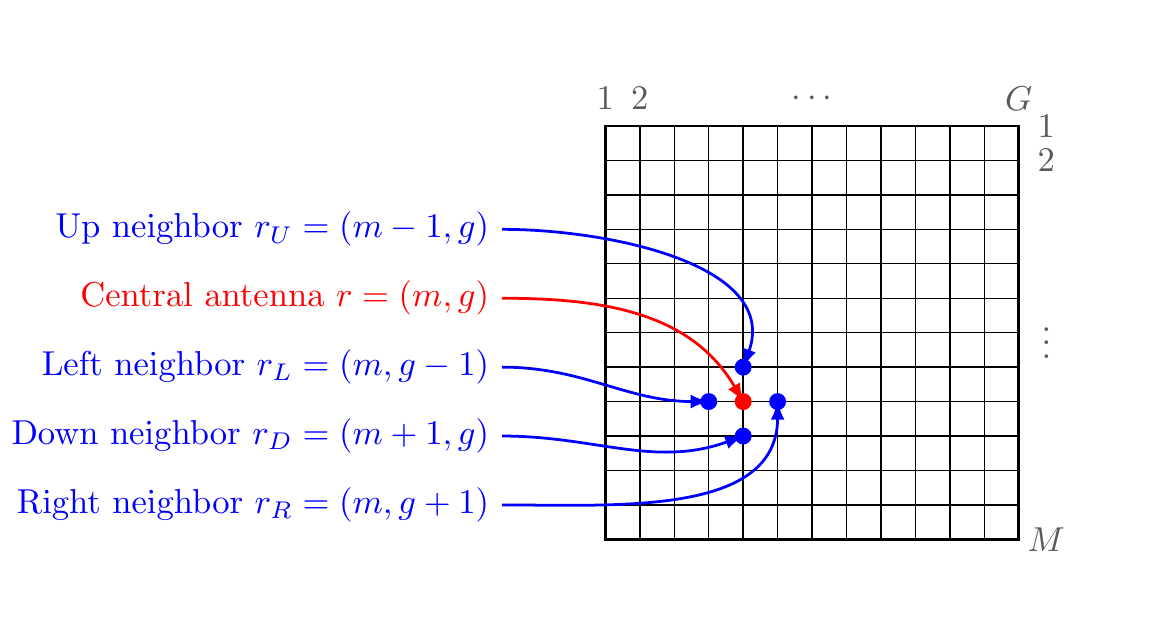}
	\caption{2-D antenna grid of size $M\times G$. An arbitrarily selected antenna is highlighted in red along with its neighboring antennas in blue. In this context, the red antenna is the central antenna $r$ and $r_U, r_R, r_D,$ and $r_L$ are its 4-neighbors.}
\label{fig:2Dantennagrid}
\end{figure}

Let $\hv^{r} \in \mathbb{C}^L$ denote the CIR which models the channel between a typical single antenna user and the receive antenna $r=(m,g)$ where $m\in\{1,2,\dots, M\}$ and $g\in\{1,2,\dots, G\}$ as shown in Fig. \ref{fig:2Dantennagrid}.
We assume that $\hv^r$ is sparse and is modeled as \cite{6280684}
\begin{align}
\hv^r = \hv_A^r \odot \hv_B^r,
\end{align}
where $\odot$ indicates element-by-element multiplication. The vector $\hv_A^r$ consists of elements that are drawn from some distribution\footnote{We put no restriction on the distribution of $\hv_A^r$ which could be Gaussian or not. The distribution might even be unknown and the coefficients of $\hv_A^r$ need not be iid. The implementation in this paper is agnostic to the distribution of channel coefficients.} and $\hv_B^r$ is a Bernoulli random vector with independent entries that are distributed as \cite{6280684}
\begin{align}\label{eq:priorshB}
 \mathrm{P}(h_B^r(i) = j) &=\begin{cases}
    \lambda_i, & \text{for $j=1$}.\\
    1-\lambda_i, & \text{for $j=0$}.
  \end{cases}
\end{align}
In other words, the entries of $\hv_B^r$ form a collection of independent (and possibly non-identically distributed) Bernoulli random variables. Thus, $\hv^r$ is an $L$-tap discrete-time sparse channel, where no assumption whatsoever is made about the distribution of its non-zero complex-valued coefficients.

The received signal at the $r$th antenna is best described in the frequency domain and is given by
\begin{align}\label{eq:received_signal_freq_domain}
\Ybc^r &= \mathrm{diag}(\bm{\Xc}) \Hbc^r + \Wbc^r,
\end{align}
where $\Ybc^r$ is obtained from the time-domain received signal by removing the cyclic prefix and pre-multiplying by the Fourier matrix $\Fm$. The noise $\Wbc^r \sim \Cc\Nc(\zerov, \sigma_w^2\Id)$ is the frequency-domain noise vector of dimension $N\times 1$ and $\Hbc^r$ is the $N \times 1$ channel frequency response vector i.e.,
\begin{align}\label{eq:Hr}
      \Hbc^r &=\Fm\begin{bmatrix}\hv^r\\
                            \zerov_{N-L \times 1}
      \end{bmatrix}=\underbar{\Fm}\hv^r
\end{align}
where $\underbar{\Fm}$ is the truncated Fourier matrix of size $N \times L$ formed by selecting the first $L$ columns of $\Fm$. Using (\ref{eq:Hr}), we can rewrite (\ref{eq:received_signal_freq_domain}) as
\begin{align}\label{eq:received_signal_freq_domain_final}
\Ybc^r &= \mathrm{diag}(\bm{\Xc}) \underbar{\Fm}\hv^r + \Wbc^r= \Am\hv^r + \Wbc^r,
\end{align}
where $\Am \triangleq \mathrm{diag}(\bm{\Xc})\underbar{\Fm}$ is an $N \times L$ matrix.

\subsection{Spatial Channel Model}\label{sec:SpatialChannelModel}

The large number of antennas in massive MIMO can be arranged in different configurations. For example, \begin{inparaenum}[\itshape a\upshape)]
\item linear,
\item planar (rectangular), and
\item cylindrical (circular).
\end{inparaenum} Our algorithm is capable of working on any configuration as will be explained later in the paper. However, for convenience, we adopt the uniform rectangular array.

In massive MIMO it is reasonable to assume that antenna elements in the same vicinity will observe almost same echoes from different scatterers and therefore the corresponding channels will have common support. For the wireless system under study, the signal bandwidth, operating frequencies and antenna separation controls the supports commonality across the large arrays.
Specifically, the time difference of arrival $\Delta \tau$ of a wavefront to two antennas separated by a distance $d$ satisfies $\Delta \tau \le \frac{d}{C}$, where $C$ is the speed of light. The authors in \cite{barbotin2011estimating} suggest that two channel taps are resolvable if the time difference of arrival is larger than $\frac{1}{10BW}$ where $BW$ is the signal bandwidth. Thus, let $d_{\mathrm{max}}$ be the distance between the farthest antennas of an array ($d_{\mathrm{max}}$ is a function of the antenna spacing $d$ and the number of antennas in the array), then it follows easily from the above that the antenna array might exhibit one of the two possible scenarios:

\subsubsection{The array is spatially invariant with respect to the CIR support if $\frac{d_{\mathrm{max}}}{C} \le \frac{1}{10BW}$}

\begin{figure}[htbp]
    \centering
    \includegraphics[width=0.75\columnwidth]{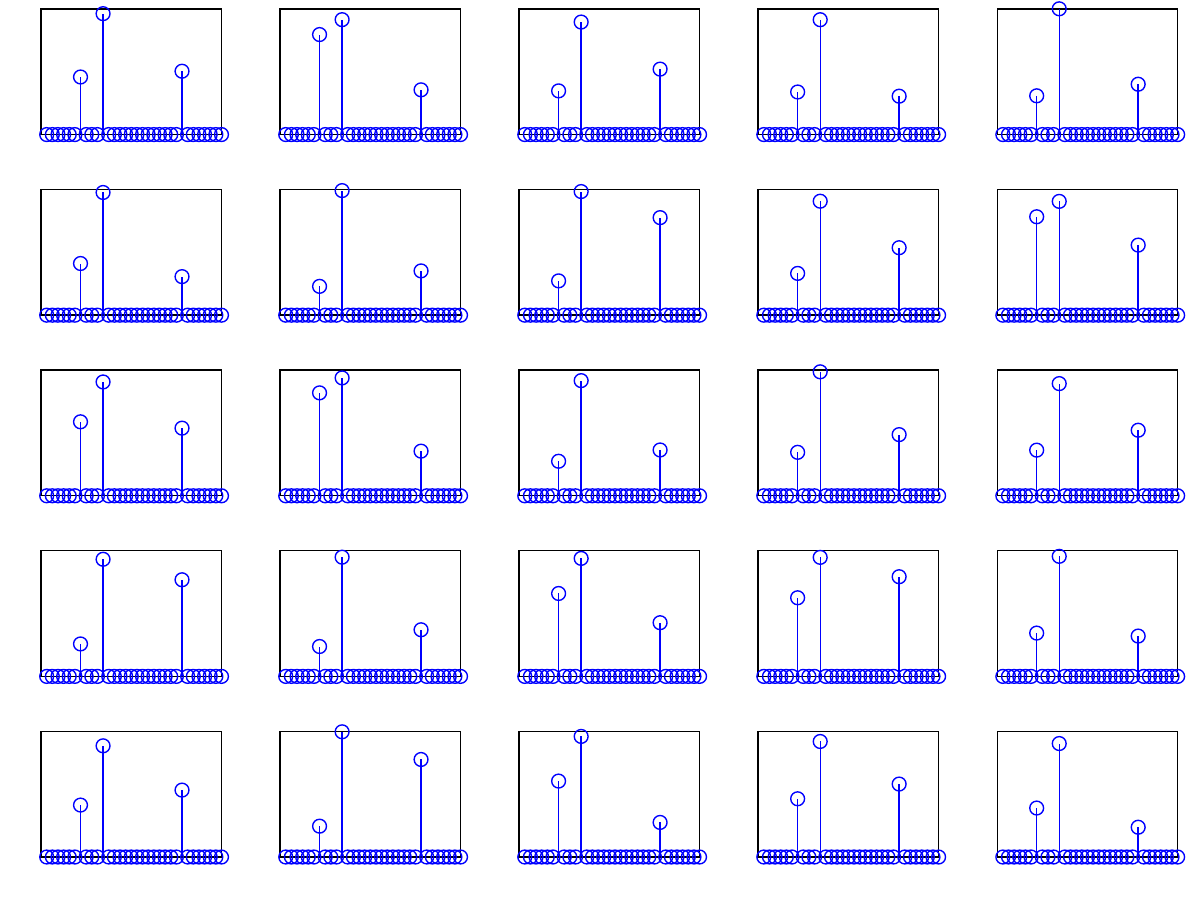}
	   \caption{CIRs of a $5\times 5$ section of a space-invariant antenna array. Plots show the tap strengths on y-axis with respect to the tap locations on x-axis for antennas in this space-invariant antenna array section. Note that the support is invariant across the array but the taps strengths fade differently across the array.}
    \label{fig:SIA}
    \end{figure}

Here, all $\hv^r$'s will have same sparsity pattern, which means the amplitudes of the channel taps might be different but the positions of the most significant taps (MST) will not change. This assumption is motivated by the fact that for closely spaced antenna elements, the times of arrival are quite close, though the paths amplitudes and phases could be different. Therefore, different antenna elements will experience almost the same echoes from the different scatterers. In other words, the support of the channels will not change as we move from one antenna to another throughout the array but the tap strengths might be different as evident from Fig. \ref{fig:SIA}. We call such arrays space-invariant arrays (SIA).

\subsubsection{The array is spatially variant with respect to the CIR support if $\frac{d_{\mathrm{max}}}{C} > \frac{1}{10BW}$}\label{sec:SpaceVariant}

In this case, the channel support varies across the array. Note that such variation takes place slowly and, therefore, it is safe to assume that the following property is always valid.%
\begin{center}
\fbox{%
  \parbox{0.9\columnwidth}{%
    \begin{center}
        \textbf{Property 2:} \emph{Any central antenna and its 4-neighbors have \underline{approximately} common support.}
    \end{center}
  }%
}
\end{center}

\begin{figure}[htbp]
    \centering
    \includegraphics[width=0.75\columnwidth]{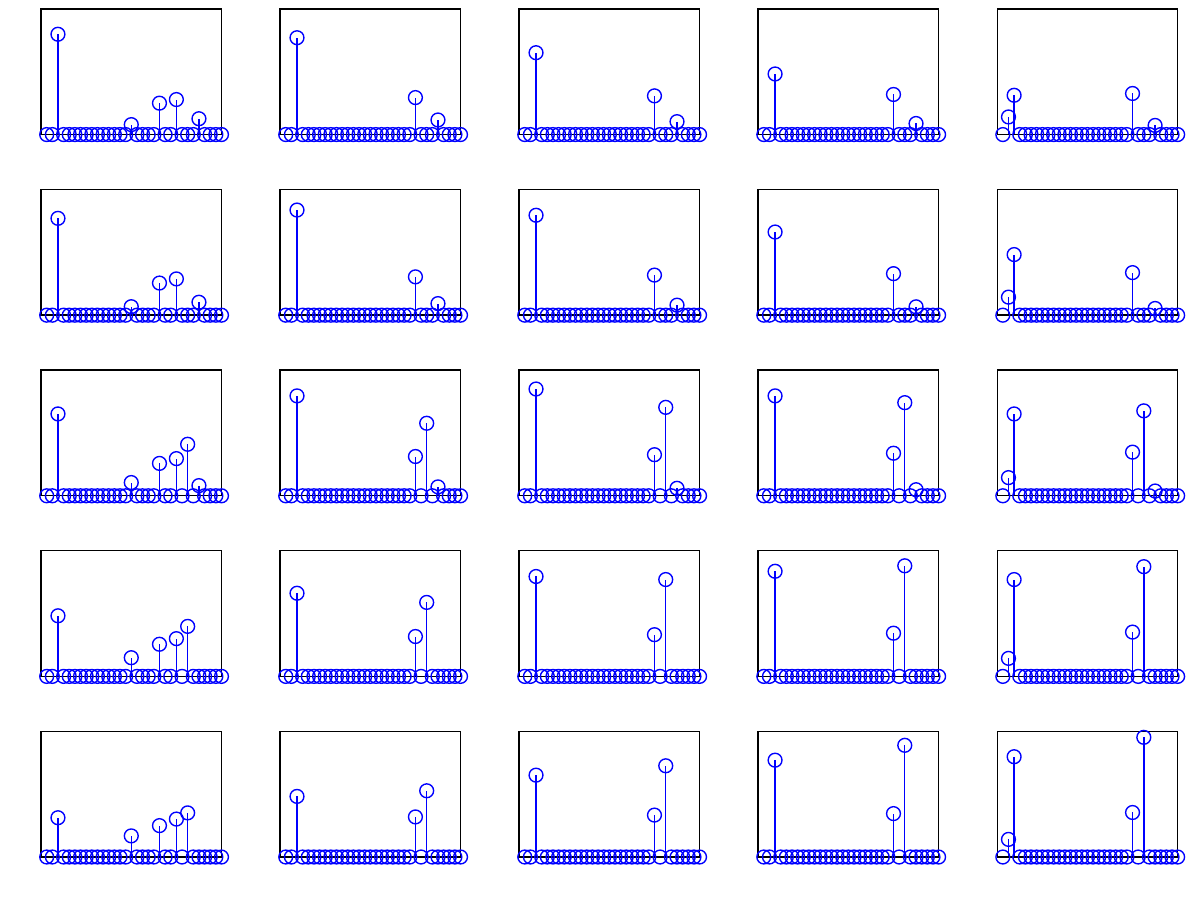}
	   \caption{CIRs of a $5\times 5$ section of a space-variant antenna array. Plots show the tap strengths on y-axis with respect to the tap locations on x-axis for antennas in this space-variant antenna array section. Note that neighboring antennas have \emph{approximately} the same support.}
    \label{fig:SVA}
    \end{figure}

See Fig. \ref{fig:SVA} for an example where the neighboring antennas have approximately the same support. We call such arrays space-variant arrays (SVA).

In Table \ref{table:system_parameters} we classify antenna arrays of three different dimensions as either SIA or SVA. Specifically, the table illustrates the relationship between the maximum resolvable distance ($d_{\mathrm{max}}$) and the dimensions of the arrays for three different communications standards. For instance, in the 3GPP LTE standard, the distance between two antenna elements on the far ends of a $10 \times 10$ array is $9d<d_{\mathrm{max}}$, thus the array is SIA. Whereas, for a $50 \times 50$ array, the distance is $49d > d_{\mathrm{max}}$ causing the array to be SVA. Note that in this table the distance between two adjacent antennas is assumed to be $d = \lambda/2$ where $\lambda$ is the signal wavelength.

While most of the available research in MIMO channel estimation deals with the space-invariant case (for example, \cite{semi_blind_MIMO_OFDM, barbotin2011estimating, blind_sparse, blind_sparse2}), very limited research has been conducted for the space-variant scenario. Similarly, the literature related to the estimation of space-variant sparse channels in massive MIMO is limited (e.g., see \cite{library977759} and the references therein). The approach we pursue in this paper is capable of dealing with both the space-variant and space-invariant cases.

\begin{table*}[htbp]
\centering
\renewcommand{\arraystretch}{1.3}
{\small
\begin{tabular}{c | c |p{2.1cm} |c |c |c |c |c}
\hline\hline
\textbf{Standard} & \textbf{Bandwidth} ($BW$)& \centering{\textbf{Center frequency } ($f_c$)} & ${d}_{\mathrm{max}}=\frac{C}{10 BW}$  & $d=\frac{\lambda}{2}$ & $10 \times 10$ array & $50 \times 50$ array & $100 \times 100$ array  \\ 
\hline \hline 
CDMA2000 & $1.25$ MHz & \centering{$1$ GHz} & $24$ m & $0.150$ m & SIA & SIA & SIA \\ 
\hline 
3GPP LTE & $20$ MHz & \centering{$2.6$ GHz} & $1.5$ m & $0.058$ m & SIA & SVA & SVA \\ 
\hline 
UWB & $500$ MHz & \centering{$3$ GHz} & $0.06$ m & $0.050$ m & SVA & SVA & SVA \\ 
\hline 
\end{tabular}}
\caption{Wireless Systems Parameters}
\label{table:system_parameters}
\end{table*}

\subsection{Pilots}

Pilots are needed for channel estimation where the transmitter reserves $K$ subcarriers for pilots and uses the remaining $N-K$ carriers for data transmission. Let $\Pc$ denote the set of indices of pilot carriers. Using (\ref{eq:received_signal_freq_domain_final}), the received pilots at receive antenna $r$ are then given by
\begin{align}\label{eq:probmodel}
\Ybc^r(\Pc) = \Am(\Pc) \hv^r + \Wbc^r(\Pc)
\end{align}
where $\Ybc^r(\Pc)$ and $\Wbc^r(\Pc)$ are $K \times 1$ vectors formed, respectively, by selecting the $K$ entries of $\Ybc^r$ and $\Wbc^r$ indexed by $\Pc$. Similarly, $\Am(\Pc)$ is a $K \times L$ matrix formed by selecting the rows of $\Am$ indexed by $\Pc$. Solving (\ref{eq:probmodel}) for $\hv^r$ obviously requires that we at least have more pilots than the channel delay spread (i.e., $K\ge L$), which impacts the spectral efficiency of the system. Here, however, we use the sparse nature of the channel and the fact that adjacent antennas have almost  the same support (i.e., properties 1 and 2) to substantially reduce the number of pilots needed as promised by the compressed sensing theory \cite{donoho,candes}.

Several pilot placement schemes have been suggested for OFDM channel estimation. It is best to allocate the pilots uniformly in conventional OFDM channel estimation (which does not make use of sparsity) \cite{5715843,5947173,1284837,1318954}. However, when the channel is sparse a random assignment of pilots has been observed to be optimal \cite{det_pilots_sparse,random_pilots_sparse2}.

With this model, we are now ready to tackle the problem of channel estimation. We do that in three steps spread over three sections
\begin{enumerate}
\item Bayesian channel estimation at each antenna,
\item Distributed channel estimation, and
\item Data-aided channel estimation.
\end{enumerate}


\section{Sparsity-aware Distribution Agnostic Bayesian Channel Estimation}\label{sec:sabmp}

Consider the linear regression model presented in (\ref{eq:probmodel}). For notational convenience, we will drop the superscript $r$ and the symbol $\Pc$ unless these are required for clarity. Hence (\ref{eq:probmodel}) becomes
\begin{align}\label{eq:sigmodel}
\Ybc = \Am \hv + \Wbc,
\end{align}
where $\Ybc$ and $\Wbc$ are vectors of dimension $K \times 1$, $\hv$ is a vector of dimension $L \times 1$ and $\Am$ is a matrix of dimension $K \times L$. Here we are interested in performing Bayesian estimation of the wireless CIR $\hv$. Bayesian approaches assume a prior distribution, however, given the dynamic nature of  wireless channels, it is usually impossible to characterize the distribution. Moreover, such an assumption is usually not suitable as it does not reflect the reality and might result in performance degradation. Additionally, even if the distribution is known, it is very difficult to estimate the distribution parameters (e.g., mean and variance for Gaussian), especially when the channel statistics are not i.i.d. In that respect, the use of distribution agnostic Bayesian sparse signal recovery (SABMP) developed by the authors in \cite{sabmp, wosspa_sabmp} is quite attractive, as it provides Bayesian estimates even when the prior is non-Gaussian or unknown.

\subsection{Simple Channel Estimation using SABMP}\label{sec:simple_chann_est}

The set of channel estimation algorithms that we propose in this paper (Sec. \ref{sec:coordinated} and \ref{sec:data-aided}) use a modified version of the \texttt{SABMP} algorithm proposed by the authors in \cite{sabmp, wosspa_sabmp}. The modifications to \texttt{SABMP} required for the development of our distributed and data-aided channel estimation methods are proposed in Sec. \ref{sec:covar} and \ref{sec:FindingMarginals}. However, before presenting the modifications we consider it essential to quickly go through the steps followed by the \texttt{SABMP} algorithm. In that respect, we briefly describe a straightforward approach for sparse channel estimation using \texttt{SABMP}. In this approach, all channels $\hv^r$ are estimated independently using the \texttt{SABMP} algorithm. Since no collaboration takes place among antennas in this approach, it is not possible to take advantage of property 2 mentioned earlier.

To estimate the $L \times 1$ sparse channel $\hv$, from the $K \times 1$ observations vector $\Ybc$ related by the linear regression model given in (\ref{eq:sigmodel}), \texttt{SABMP} pursues an MMSE estimate of $\hv$ given $\Ybc$ which is formally defined by
\begin{equation}\label{eq:app:xmmse}
\widehat{\hv}_{\rm{MMSE}} \triangleq \mathbb{E}[\hv|\Ybc] = \sum_{\Sc} p(\Sc|\Ybc)\mathbb{E}[\hv|\Ybc,\Sc].
\end{equation}
Here the sum is executed over all possible $2^{L}$ support sets of $\hv$. However, computing this sum is a challenging task when the channel delay spread $(L)$ is large because the number of possible support sets can be extremely large and the computational complexity will become unrealistic.  To have a computationally feasible solution, this sum can be approximated by considering only those support sets which include the most significant taps with high probability. These few support sets correspond to the sets with significant posteriors $p(\Sc|\Ybc)$. Let $\Sc_d$ be the set of supports for which the posteriors are significant. Hence, (\ref{eq:app:xmmse}) can be approximated by\footnote{Note that $\sum_{\Sc\in\Sc_d} p(\Sc|\yv) < 1$ since $\Sc_d \subset \Sc$. This would render the estimate in (\ref{eq:app:xammse}) biased. To ensure an unbiased estimate, we normalize $p(\Sc|\yv)$ so that $\sum_{\Sc\in\Sc_d} p(\Sc|\yv) = 1$.}
\begin{equation}\label{eq:app:xammse}
\widehat{\hv}_{\rm{AMMSE}} = \sum_{\Sc\in\Sc_d} p(\Sc|\Ybc)\mathbb{E}[\hv|\Ybc,\Sc].
\end{equation}
We could determine $\Sc_d$ and $\widehat{\hv}_{\rm AMMSE}$ in a greedy manner using the dominant support selection metric defined as the log posterior
\begin{align}\label{eq:app:dssm1}
\nu(\Sc) &\triangleq \ln p(\Sc | \Ybc) = \ln p(\Ybc|\Sc) p(\Sc).
\end{align}

The greedy algorithm of \texttt{SABMP} starts by first finding the best support of size 1. This requires evaluating $\nu(\Sc)$ for $\Sc=\{ 1 \}, \dots, \{ L \}$, i.e., a total of $\binom{L}{1}$ search points. Let $\Sc_1 = \{ \alpha_1 \}$ be the optimal support. Now, the optimal support of size 2 is found. Ideally, this involves a search over a space of size $\binom{L}{2}$. To reduce the search space, however, the greedy approach looks for the tap location $\alpha_2 \neq \alpha_1$ such that $\Sc_2=\{ \alpha_1, \alpha_2 \}$ maximizes $\nu(\Sc_2)$. This involves $\binom{L-1}{1}$ search points (as opposed to the optimal search over $\binom{L}{2}$ points). The process continues in this manner by forming $\Sc_3 = \{ \alpha_1, \alpha_2, \alpha_3 \}$ and so on. Therefore, $\Sc_d$, the set of dominant support sets is composed of support sets that are incremental in nature and is given by\footnote{In (\ref{eq:app:Sd}), $T_{\mathrm{max}}$ refers to the maximum number of non-zero elements in the sparse $\hv$. $T_{\mathrm{max}}$ is selected to be slightly larger than the expected number of active taps in the estimated CIR using the de Moivre-Laplace theorem. For details, readers are referred to \cite{sabmp}.\label{ftn:Tmax}}
\begin{align}\label{eq:app:Sd}
\Sc_d &= \left\{ \Sc_1, \Sc_2, \dots, \Sc_{T_{\mathrm{max}}}  \right\},\nonumber\\
\Sc_d &= \left\{  \{ \alpha_1\}, \{ \alpha_1, \alpha_2\}, \{ \alpha_1, \alpha_2, \alpha_3\}, \dots, \{ \alpha_1, \alpha_2, \dots, \alpha_{T_{\mathrm{max}}}\} \right\}.
\end{align}

The development of the \texttt{SABMP} algorithm in \cite{sabmp} assumes that the taps of $\hv$ are activated with equal probability $\lambda$ (i.e., i.i.d. Bernoulli with probability $\lambda$). However, here we consider the case where some taps are more probable than others (based on the available information), and hence it is desirable to assign those taps a higher probability. This requires us to assume an independent and non-identically distributed Bernoulli behavior for the unknown sparse vector and therefore the prior is given by
\begin{align}\label{eq:app:pS}
p(\Sc) &=  \prod_{i\in \Sc} \lambda_i \prod_{j\in\{1,\dots,L\}\backslash \Sc } (1-\lambda_j),
\end{align}
where, $\lambda_i$ is the probability that the $i$th tap of $\hv$ is active. Moreover, the likelihood is approximated as
\begin{align}\label{eq:app:pyS}
p(\Ybc|\Sc) &=  \exp \left( -\frac{1}{2\sigma_w^2} \left\|\Pm_\Sc^{\bot} \Ybc\right\|_2^2 \right),
\end{align}
where, $\Pm_\Sc^\bot =\Id - \Pm_\Sc  =\Id - \Am_\Sc\left( \Am_\Sc^\herm \Am_\Sc\right)^{-1} \Am_\Sc^\herm$ is the projection matrix and $\Am_\Sc$ is a matrix formed by selecting columns of $\Am$ indexed by support $\Sc$. Substituting (\ref{eq:app:pS}) and (\ref{eq:app:pyS}) in (\ref{eq:app:dssm1}) yields
\begin{align}\label{eq:app:dssm2}
\nu(\Sc) \triangleq \ln p(\Sc | \Ybc) &= ( -\frac{1}{2\sigma_w^2}) \left\|\Pm_\Sc^{\bot} \Ybc\right\|_2^2 + \sum_{i\in \Sc}\ln \lambda_i \nonumber\\
&+ \sum_{j \in \{1,\cdots,L\}\backslash\Sc}\ln(1-\lambda_j)
\end{align}
Now the only term that is left to be evaluated in (\ref{eq:app:xammse}) is  $\mathbb{E}[\hv|\Ybc,\Sc]$. Note that it is difficult or even impossible to evaluate this quantity because the distribution of the active taps of $\hv$ is unknown. Therefore, we replace it by the best linear unbiased (BLUE) estimate given by
\begin{align}\label{eq:BLUE}
\mathbb{E}[\hv|\Ybc,\Sc] \leftarrow \left( \Am_\Sc^\herm \Am_\Sc\right)^{-1} \Am_\Sc^\herm \Ybc.
\end{align}
This provides us all the required quantities to evaluate $\widehat{\hv}_{\rm AMMSE}$. Note that all parameters including $\sigma_w^2, \bm{\lambda}=\{\lambda_i\}_{i=1}^L$ and the possible size of support $T_{\rm max}$ need not be known and are estimated by the algorithm. Specifically, $\lambda_i$'s are initialized as
\begin{align*}
\lambda_{i} &= \frac{1}{L}\left|  \left\{  j :  \left| \av_j^\herm \Ybc \right| \ge \frac{1}{2} \left\| \av^\herm \Ybc \right\|_\infty \right\}  \right|,
\end{align*}
where $\av_j$ is the $j$th column of the matrix $\Am$. Moreover, $\sigma_w^2$ is initialized simply as a scaled version of the variance of the received signal i.e., $\sigma_\Ybc^2$. Finally, $T_{\mathrm{max}}$ is selected to be slightly larger than the expected number of active taps in the estimated CIR using the de Moivre-Laplace theorem. Note that our algorithm is robust to these initial estimates and can find right support even if these parameters are initialized away from their true values. For more details the interested readers are referred to \cite{sabmp}.

By following this greedy approach, each antenna node estimates the corresponding approximate sparse CIR (\ref{eq:app:xammse}) in a distribution agnostic manner. A detailed statement of the greedy algorithm is presented in Table \ref{alg:greedy}.

\begin{table}
\begin{algorithmic}[1]
\Procedure{Greedy} {$\Am, \Ybc, \bm{\lambda}, \sigma_w^2, T_{\rm max}$} 
\State \textbf{initialize} $J \gets \{1, 2, \hdots, L\},\, i \gets 1$
\State \textbf{initialize empty sets} $\Sc_{max},\, \Sc_d,\, p(\Sc_d|\Ybc),\, \mathbb{E}[\hv|\Ybc,\Sc_d]$
\State $J_i \gets J$
\While{$i \le T_{\rm max}$} 
\State $\Omega \gets  \{ \Sc_{max} \cup \{ \alpha_1 \},
 \Sc_{max} \cup \{ \alpha_2 \},
\cdots,
 \Sc_{max} \cup \{ \alpha_{|J_i|} \} \mid \alpha_k \in J_i\}$
\State \textbf{compute }$\{ \nu(\Sc_k) \mid \Sc_k \in \Omega \}$
\State \textbf{find} $\Sc_\star \in \Omega$ \textbf{such that} $\nu(\Sc_\star) \ge \max_j \nu(\Sc_j)$
\State $\Sc_d \gets \{\Sc_d, \Sc_\star\}$
\State \textbf{compute} $p(\Sc_\star|\Ybc), \mathbb{E}[\hv|\Ybc,\Sc_\star]$
\State $p(\Sc_d|\Ybc) \gets \{p(\Sc_d|\Ybc), p(\Sc_\star|\Ybc)\}$
\State $\mathbb{E}[\hv|\Ybc,\Sc_d] \gets \{\mathbb{E}[\hv|\Ybc,\Sc_d] , \mathbb{E}[\hv|\Ybc,\Sc_\star\}$
\State $\Sc_{max}\gets \Sc_\star$
\State $J_{i+1} \gets L~\backslash~\Sc_\star$
\State $i \gets i+1$
\EndWhile\label{euclidendwhile}
\State\label{alg:greedy:returnline} \textbf{return} $\Sc_d, p(\Sc_d|\Ybc), \mathbb{E}[\hv|\Ybc,\Sc_d]$
\EndProcedure
\end{algorithmic}
\caption{Support Agnostic Bayesian Matching Pursuit Algorithm (SABMP)}\label{alg:greedy}
\end{table}

In addition to the non-iid generalization above, we develop in the following two necessary modifications to the \texttt{SABMP} algorithm. Specifically, we modify \texttt{SABMP} to output \begin{inparaenum}[\itshape a\upshape)]
\item the channel estimation error covariance matrix and
\item the marginal probabilities of the detected MSTs
\end{inparaenum} that are needed for the distributed and data-aided versions of the channel estimation algorithms proposed in Sec. \ref{sec:coordinated} and \ref{sec:data-aided} respectively.

\subsection{Error Covariance and Estimation Error}\label{sec:covar}

The channel estimation error and the covariance could be computed as follows.

Let,
\begin{align}\label{eq:channelerror}
\widetilde{\hv} = \widehat{\hv}_{\rm AMMSE} - \hv
\end{align}
be the error vector and  $\Rm_{\widetilde{\hv}} \triangleq {\rm cov}[\widetilde{\hv}|\Ybc]$ represent the error covariance matrix. The trace of $\Rm_{\widetilde{\hv}}$ i.e., $\Tr[\Rm_{\widetilde{\hv}}]$ gives the MMSE estimation error. In order to evaluate $\Rm_{\widetilde{\hv}}$, let us define the error vector $\widetilde{\hv}_\Sc = \widehat{\hv}_\Sc - \hv$ for a given support $\Sc$, where $\widehat{\hv}_\Sc = \mathbb{E}[\hv|\Ybc,\Sc]$. Let the corresponding error covariance matrix be $\Rm_{\widetilde{\hv}|\Sc} \triangleq {\rm cov}[\widetilde{\hv}|\Ybc,\Sc]$. Then $\Rm_{\widetilde{\hv}}$ could be expressed in terms of $\Rm_{\widetilde{\hv}|\Sc}$ by summing it over the dominant support set $\Sc_d$ and is given by
\begin{align}\label{eq:Rh}
\Rm_{\widetilde{\hv}} &= \sum_{\Sc\in\Sc_d} p(\Sc|\Ybc) \,\, \Rm_{\widetilde{\hv}|\Sc}.
\end{align}
Since we replace $\mathbb{E}[\hv|\Ybc,\Sc]$ with a BLUE estimate, the conditional error covariance matrix will be $\Rm_{\widetilde{\hv}|\Sc} = (\Am_\Sc^\herm \Cm^{-1} \Am_\Sc)^{-1}$ \cite{poor1994introduction} (where $\Cm = \sigma_w^2 \Id$ is the noise covariance matrix). Combining this fact with (\ref{eq:Rh}) yields
\begin{align}\label{eq:error_covariance_matrix}
\Rm_{\widetilde{\hv}} &= \sigma_w^2\sum_{\Sc\in\Sc_d} p(\Sc|\Ybc) \,\, (\Am_\Sc^\herm \Am_\Sc)^{-1}.
\end{align}
Note that the calculation of covariance matrix involves a matrix inversion term which is a computationally expensive task. However, we would like to highlight that these inverses are available as part of intermediate calculations in the \texttt{SABMP} algorithm and hence do not pose any additional computational burden. The error covariance matrix and the estimation error play a vital role in the development of the data-aided approach presented in Sec. \ref{sec:data-aided}.

\subsection{Finding Marginals}\label{sec:FindingMarginals}
The marginal probabilities are not directly available at the output of \texttt{SABMP} and could be computed from the posteriors $p(\Sc|\Ybc), \forall \Sc\in \Sc_d$ in a simple manner described below.

Let $T^r = \{  \alpha_1^r, \alpha_2^r, \dots, \alpha_{T_{\rm max}}^r\}$ be the set of MST locations of channel $\hv^r$ as detected by \texttt{SABMP} algorithm. Then the marginal probabilities  of $\alpha_i^r$,  $\forall i\in \{1,2,\dots,T_{\rm max} \}$ could be computed as
\begin{align}\label{eq:palphat}
p(\alpha_i^r|\Ybc) &= \sum_{\alpha_i^r \cap \Sc \ne \varnothing } p(\Sc|\Ybc),
\end{align}
where the sum is evaluated over all $\Sc \in \Sc_L^r$ satisfying the condition ${\alpha_i^r \cap \Sc \ne \varnothing }$ where $\Sc_L^r$ contains all $2^L - 1$ support sets that could be created for $\hv^r$ (recall that $\hv^r$ is a vector of length $L$). Among these $2^{L}-1$ support sets there are only $2^{T_{\rm max}}-1$ support sets which involve purely the ${T_{\rm max}}$ detected non-zero locations. Let us denote the set of these $2^{T_{\rm max}}-1$ support sets by $\Sc_{T_{\rm max}}^r$. We assert that only the support sets present in $\Sc_{T_{\rm max}}^r$ have significant posteriors $p(\Sc|\Ybc)$ as compared to the others which have very small values. This follows from the findings of \cite{sabmp} (see Fig. 7 therein). Thus, we can evaluate the sum in (\ref{eq:palphat}) over $\Sc \in \Sc_{T_{\rm max}}^r$ to find the marginal probability of each detected non-zero location. However, the \texttt{SABMP} algorithm returns $\Sc_d^r$, which unlike $\Sc_{T_{\rm max}}^r$ contains only ${T_{\rm max}}$ support sets (see (\ref{eq:app:Sd})). Therefore, we modify the \texttt{SABMP} algorithm so that it outputs $p(\Sc|\Ybc)$ for all supports in $\Sc_{T_{\rm max}}^r$. For illustration purpose Table \ref{tab:Sd_vs_Sp} provides an example of the support sets $\Sc_d^r$ and $\Sc_{T_{\rm max}}^r$ when ${T_{\rm max}}=3$. Please note that for convenience of notation, we shall from now on use $\lambda(\alpha^r_i)$ as a shorthand notation for $p(\alpha_i^r|\Ybc)$.

\begin{table}
\centering
\begin{tabular}{|c|c|c|}
  \hline
    & $\Sc_d^r$ & $\Sc_{T_{\rm max}}^r$\\
    \hline
  1 & $\{\alpha_1\}$ & $\{\alpha_1\}$ \\
  2 & $\{\alpha_1, \alpha_2\}$ & $\{\alpha_2\}$ \\
  3 & $\{\alpha_1, \alpha_2, \alpha_3\}$ & $\{\alpha_3\}$ \\
  4 &  & $\{\alpha_1, \alpha_2\}$ \\
  5 &  & $\{\alpha_1, \alpha_3\}$  \\
  6 &  & $\{\alpha_2, \alpha_3\}$  \\
  7 &  & $\{\alpha_1, \alpha_2, \alpha_3\}$\\
  \hline
\end{tabular}
\caption{Sets $\Sc_d^r$ and $\Sc_{T_{\rm max}}^r$ for ${T_{\rm max}}=3$.}
\label{tab:Sd_vs_Sp}
\end{table}

Since $\Sc_{T_{\rm max}}^r$ has more support sets, this modification obviously results in increased computational complexity. However, utilizing the available intermediate information in \texttt{SABMP} helps to compute the marginalized posterior probabilities $p(\Sc|\Ybc)$ in an efficient manner. Specifically, note that for the example of Table \ref{tab:Sd_vs_Sp}, we require posteriors of $\{\alpha_2\}, \{\alpha_3\}, \{\alpha_1,\alpha_3\},$ and $\{\alpha_2, \alpha_3\}$ in addition to those returned by \texttt{SABMP}. However, it follows from the explanation given in Sec. \ref{sec:simple_chann_est}, that the posteriors for $\{\alpha_2\}, \{\alpha_3\}$, and $\{\alpha_1,\alpha_3\}$ are already available to the algorithm by virtue of the intermediate computations. Therefore, the only missing computation which has to be performed additionally is that of $\{\alpha_2, \alpha_3\}$. The same reasoning applies for larger support sizes. Therefore, we state that the increase in computational complexity is not significant.

For ease of reference, we name the modified version of \texttt{SABMP} as \texttt{RS1}. It has the following additional features:
\begin{itemize}
 \item it considers the non-zero taps to follow independent and non-identically distributed Bernoulli behavior (as highlighted in Sec. \ref{sec:simple_chann_est})
 \item it returns the error covariance of our estimate which is needed for the data-aided part   (Sec. \ref{sec:covar})
 \item it outputs the belief/probability that a given tap is active  (Sec. \ref{sec:FindingMarginals}).
\end{itemize}
This algorithm along with \texttt{SABMP} will be used in the discussion that follows to develop the coordinated channel recovery algorithms.


\section{Coordinated Channel Estimation}\label{sec:coordinated}

In the coordinated channel recovery method, the receive antennas collaborate with each other to take advantage of property 2 and estimate the MST locations jointly. In order to realize this coordinated method, we assume baseband processing at each receive antenna with an additional processor on each baseband card to implement the collaboration strategy described in this section. At the heart of the collaboration strategy followed by the proposed method is the following simple information-sharing step.
\begin{center}
\fbox{%
  \parbox{0.9\columnwidth}{%
    \begin{center}
        \textbf{Sharing Step: } Each antenna acting as a central antenna $r_C$ receives information from its direct 4-neighbors $\Nc = \{ r_U, r_D, r_R, r_L\}$.\footnotemark
    \end{center}
  }%
}
\end{center}
\footnotetext{For the elements lying at the edges of the array the number of neighbors are different. We use $\Nc$ to denote the set of neighbors irrespective of the position of $r$ and therefore $2 \le |\Nc| \le 4 $.}

It is obvious that repetitive application of this sharing step would result in information diffusion throughout the antenna grid. For example, in the first iteration $r_C$ receives information from just the first tier of antennas (i.e., the neighboring 4 antennas). In just two iterations $r_C$ receives information from $12$ antennas (i.e., the first and second antenna tiers). Therefore, with the help of this simple step each antenna is able to incorporate information from its neighbors to enhance its decision about the MSTs of its channel. This ultimately helps in estimating the channels accurately. There are two advantages of this stepwise collaboration mechanism:
 \begin{enumerate}
 \item It gives us the flexibility to control the number of collaborators for each receiver which is essential for the space-variant case (Sec. \ref{sec:SpaceVariant}).
 \item The collaboration mechanism is computationally efficient as the antennas do not all collaborate with each other at the same time. This also reduces the communication overhead.
 \end{enumerate}

We now present two algorithms for channel estimation based on this simple stepwise information sharing strategy.

\subsection{Algorithm 1: Marginal-based Channel Estimation using Pilots}\label{sec:alg1}

We seek to solve the problem mentioned in (\ref{eq:probmodel}). The proposed algorithm  starts by estimating the sparse channels $\hv^r$ at each receive antenna $r$ using the \texttt{RS1} algorithm. We initialize the algorithm by assuming that all taps of $\hv^r$ have equal active probability $\lambda_{\mathrm{init}}$, i.e.,
\begin{align}
\mathrm{P}\left(h_{B}^r(l)= 1\right)= \lambda_{\mathrm{init}}, \quad l \in \left\{1,2, \cdots, L\right\}, \forall r.
\end{align}
Let $T^{r} = \{  \alpha_1^{r}, \alpha_2^{r}, \cdots, \alpha_{T_{\mathrm{max}}}^{r}\}$ be the set of MSTs of channel $\hv^r$ as detected by \texttt{RS1}. Here $\alpha_i^r$ is the location of the $i$th detected tap of receiver $r$. Note that since $\lambda_{\mathrm{init}}$ is same throughout the array, the number of detected MSTs, i.e.,  $T_{\mathrm{max}}$, will also be same for all receivers in the array.\footnote{The value of $T_{\mathrm{max}}$ is selected to be slightly larger than the expected number of active MSTs in the estimated CIR using the de Moivre-Laplace theorem which relies on $\lambda_{\mathrm{init}}$.} In other words, the cardinality $|T^r| = T_{\mathrm{max}}, \forall r$. However, the actual tap locations might differ from one antenna to another. Therefore, it is not necessary that $|T^{r_1} \cap T^{r_2}| = T_{\rm {max}}, \text{ for } r_1 \ne r_2$.  Along with the MSTs, \texttt{RS1} also returns the marginals $\lambda(\alpha^r_t) \triangleq \mathrm{P}\left(h_{B}^r(\alpha_t^r)=1\right), t\in\{ 1,2,\cdots, T_{\mathrm{max}}\}$. At this point, we invoke the sharing step mentioned previously and share the marginals. Hence, each antenna, acting as central antenna $r_C$, collects these marginals from its 4-neighbors and computes the average marginal for each tap given by
\begin{align}\label{eq:newmarginals}
\lambda(\alpha_i^{r_C}) &=  \begin{cases}
    \sum\limits_{j \in \Nc^+} \lambda(\alpha_i^j) \Big/ |\Nc^+|, & \text{if $\alpha_i^{r_C} \in \mathop\bigcup\limits_{j \in \Nc^+}T^j$}\\
    \lambda_{\mathrm{small}}, & \text{otherwise}
  \end{cases},
\end{align}
where $\Nc^+ =\Nc  \cup r_C$, $i\in \{1,2,\cdots,L\}$ and $\lambda(\alpha_i^{r_C})$ can be seen as the updated marginal of the $i$th tap detected at $r_C$. Here, $\lambda_{\mathrm{small}}$ is an arbitrarily small value assigned to those taps which have not been detected by any of the receivers in $\Nc^+$; it is highly probable that these taps have almost zero gains. Note that (\ref{eq:newmarginals}) is performed at each antenna as each antenna is the center of some neighbors. Moreover, each antenna repeats these sharing and averaging steps $D$ times where $D$ is selected based on whether the array under consideration is classified as SIA or SVA. This repetition allows each antenna to utilize the observations of distant antenna tiers to bolster its support estimates. Note that when $D=1$, information from only the immediate neighbors is taken into consideration and for $D=2$, information belonging to the neighbors of neighbors is also incorporated in the computations. Therefore, in this fashion, higher values of $D$ make it possible to extend the scope of information sharing to distant antennas.

In the space-invariant array case, since the MST locations do not vary across the array, contribution from as many antennas as possible will always strengthen our belief in these locations. Therefore, we may select $D=\max(M,G)$ which equals to the largest dimension of the antenna array. This particular choice of $D$ ensures that each antenna receives information from every other antenna in the array. However, one might not need to select such high value of $D$ and a smaller number of iterations might be sufficient based on the problem parameters such as the observation size ($K$) and the sparsity ($n$) of the channels. In fact we could establish a loose lower bound on $D$ in the noise free case as a function of these quantities using lemma 1 in \cite{1453780}. According to this lemma, if observations from $q$ antennas are used to recover $n$-sparse channel vectors using $K$ pilots then for a unique solution the following relationship holds
\begin{align}\label{eq:nbound}
n \le \lceil (K+q)/2\rceil -1,
\end{align}
where $\lceil \cdot \rceil$ denotes the ceiling operation. This yields the lower bound on $q$ which is given by
\begin{align}\label{eq:qbound}
q > 2n - K.
\end{align}
Furthermore, it could be easily deduced that the total number of antennas that take part in the $D$th sharing and averaging step is $2D(D+1)+1$. Thus relating this number with $q$ above we conclude that, to guarantee a unique solution in the noise free case, $D$ must satisfy
\begin{align}\label{eq:lowerboundonD_pre}
2D(D+1)+1 > 2n - K,
\end{align}
which simplifies to
\begin{align}\label{eq:lowerboundonD}
D > \sqrt{n-\frac{K}{2}-\frac{1}{4}} - \frac{1}{2}.
\end{align}
For a detailed account of the lemma and its requirements please refer to \cite{1453780}.

In the space-variant case, depending upon how fast the MST locations (support) change across the array, we might or might not gain from sharing. Specifically, if the change in support is fast, using higher values of $D$ would degrade the estimates. On the other hand, if the support changes very slowly, we expect that the neighborhood around a given antenna would behave \emph{approximately} as SIA. Therefore, we select a value for $D$ such that collaboration among antennas in that neighborhood would improve the estimates. In fact, from the discussion in Sec. \ref{sec:SpatialChannelModel}, we can determine the value of $D$ which will ensure that all sharing and averaging takes place among antennas having same support. Specifically, the number of tiers (also $D$) selected for sharing could be represented in terms of the distance between two adjacent antennas $d$ and the signal bandwidth $BW$. Note that the distance between the farthest antennas in tier 1 (i.e., when $D=1$) is $2d$. Similarly, for tier 2 this distance is $4d$. In general, the distance is directly related to $D$ and is given by $2Dd$. Now to ensure space-invariance for antennas up to tier $D$, it follows that we should require $2Dd \le \frac{C}{10BW}$ (see Sec. \ref{sec:SpatialChannelModel}). Therefore,
\begin{align*}
D \le& \frac{C}{20\cdot d\cdot BW},
\end{align*}
or
\begin{align}\label{eq:Dval_for_spaceinvariance}
D =& \left\lfloor \frac{C}{20\cdot d\cdot BW} \right\rfloor,
\end{align}
where $\lfloor \cdot \rfloor$ denotes the floor operation. The value of $D$ in (\ref{eq:Dval_for_spaceinvariance}) will ensure that sharing happens among antennas whose support is approximately the same. That said the number of pilots should also be large enough such that (\ref{eq:lowerboundonD}) is also satisfied.

\begin{figure*}
        \centering
        \begin{subfigure}[b]{0.35\textwidth}
                \includegraphics[width=\textwidth]{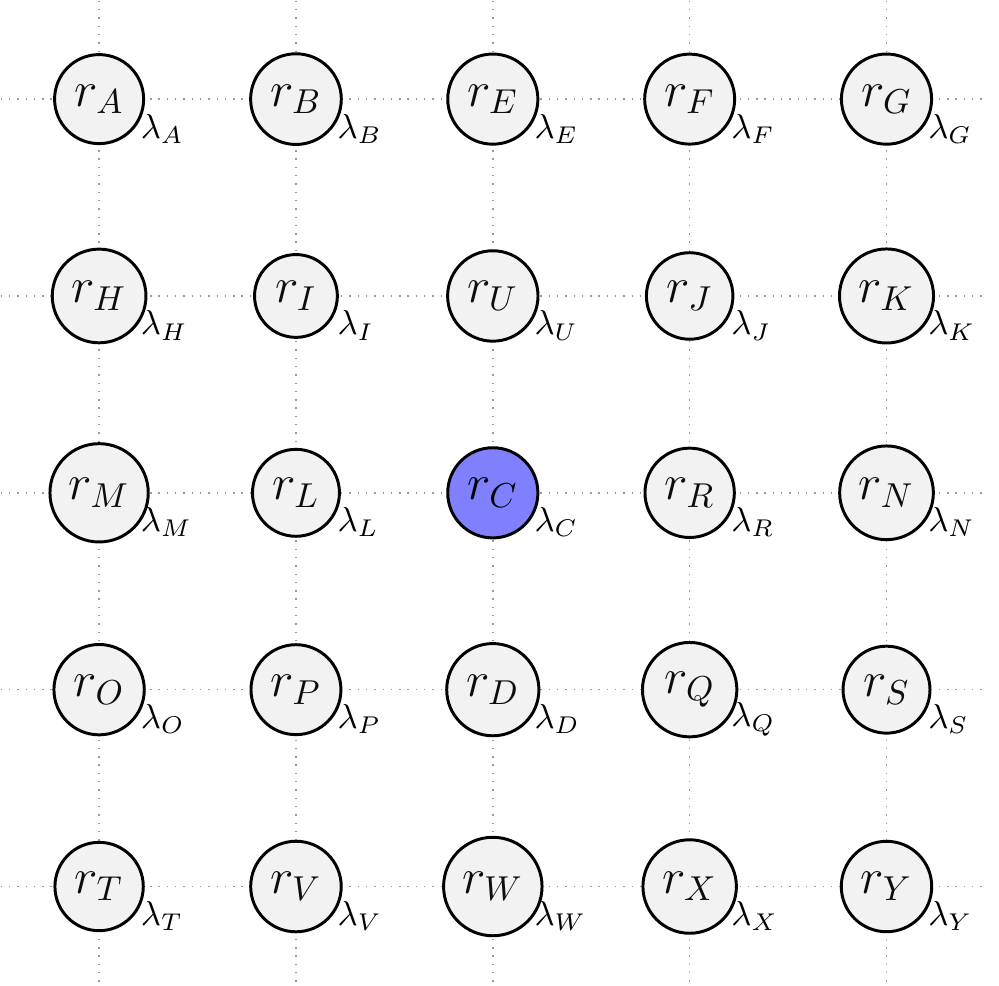}
                \caption{Step 1: Each antenna finds MSTs and their corresponding marginals $\bm{\lambda}$.}
                \label{fig:AlgStep1}
        \end{subfigure}\quad%
        ~ 
        \begin{subfigure}[b]{0.35\textwidth}
                \includegraphics[width=\textwidth]{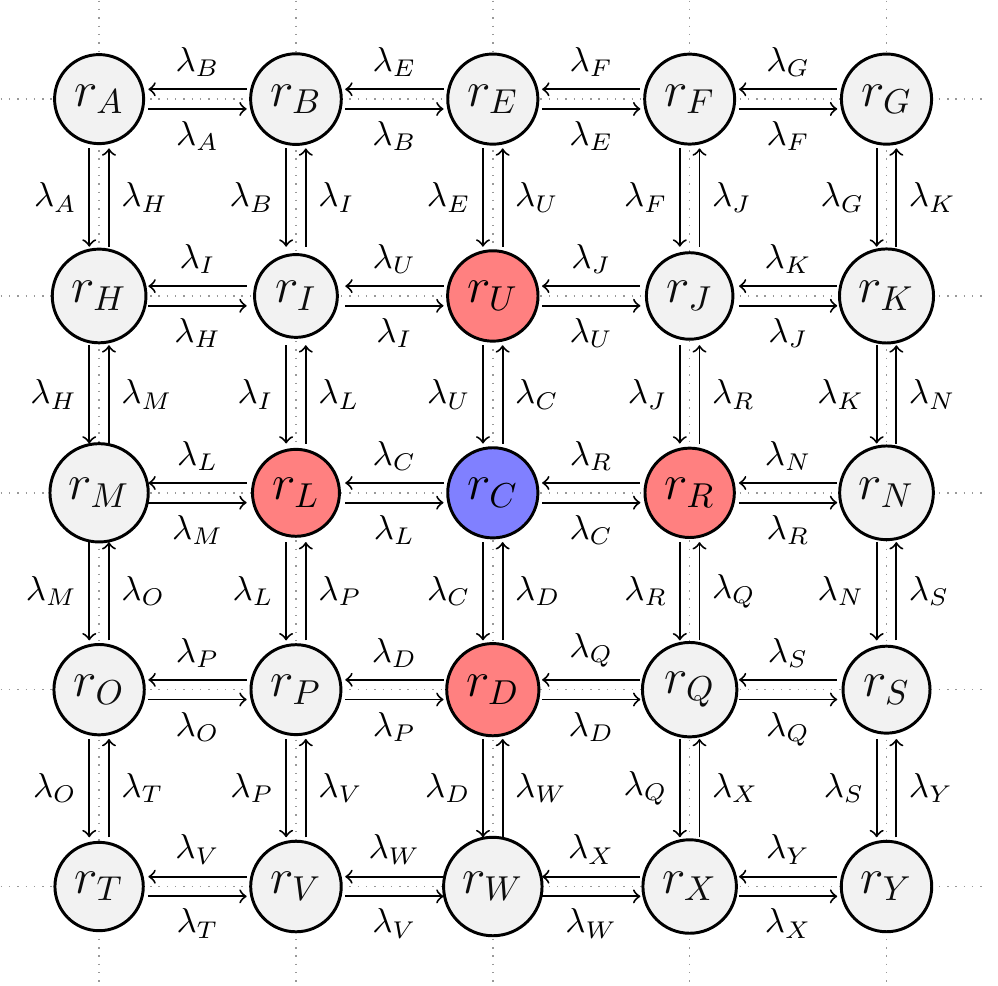}
                \caption{Step 2: Each antenna receives marginals from its 4-neighbors (highlighted red).}
                \label{fig:AlgStep2}
        \end{subfigure}\quad%
        \begin{subfigure}[b]{0.35\textwidth}
                \includegraphics[width=\textwidth]{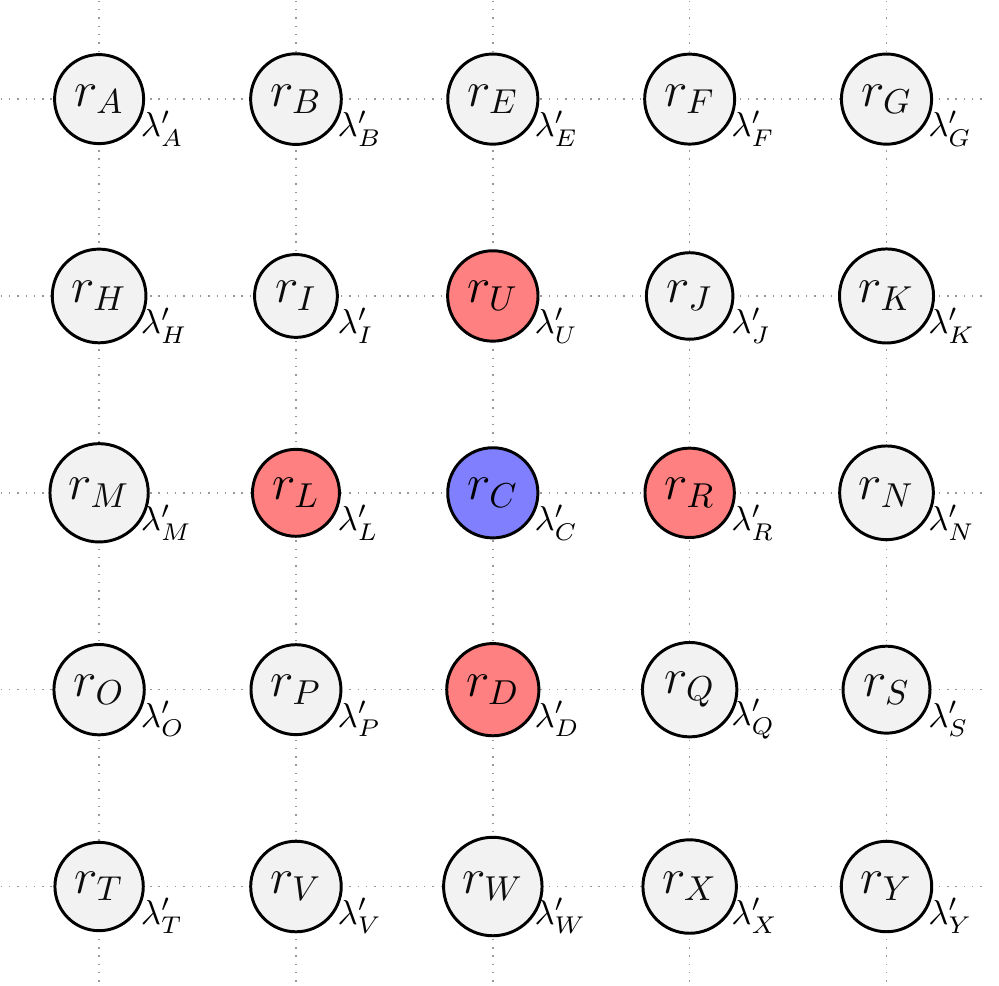}
                \caption{Step 3: Each antenna computes the mean of the received marginals $\bm{\lambda'}$.} 
                \label{fig:AlgStep3}
        \end{subfigure}\quad%
        \begin{subfigure}[b]{0.35\textwidth}
                \includegraphics[width=\textwidth]{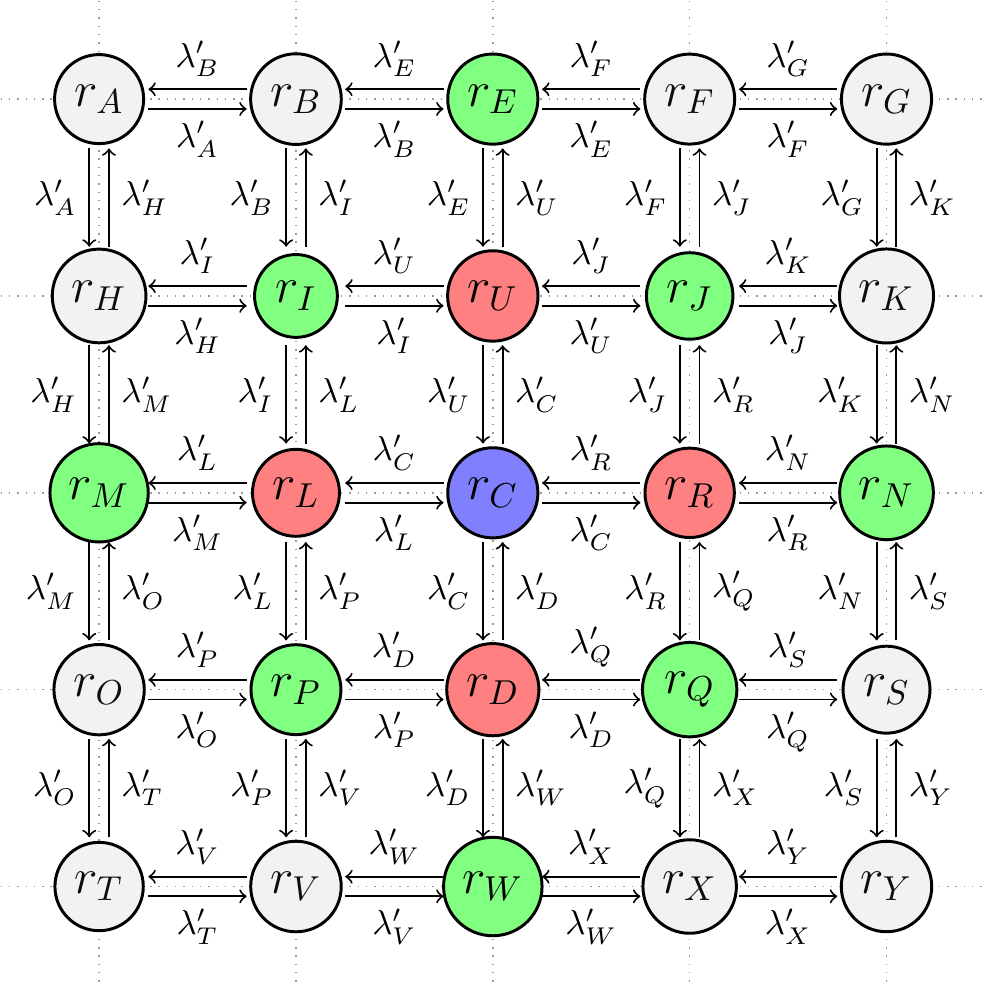}
                \caption{Step 4: Repeat steps 2 \& 3. Information from green antennas comes in.}
                \label{fig:AlgStep4}
        \end{subfigure}

        \caption{Description of the steps followed by Algorithm 1 when $D=2$. Although these steps are followed by all antennas in parallel, the process is highlighted only for the blue antenna.}\label{fig:AlgorithmSteps}
\end{figure*}

At the end of $D$ iterations each antenna has a new set of marginals  which are used as new priors with \texttt{SABMP} to get the final sparse CIR estimate. This final estimate is more accurate as the antennas have shared their information to strengthen their beliefs about the locations of the active taps.  We call this algorithm the \emph{Marginal-based Algorithm}. A graphical description of the algorithm is given in Fig. \ref{fig:AlgorithmSteps} and a summary of the steps followed by the algorithm is presented in Algorithm \ref{alg:Alg1}.

\begin{algorithm}
\caption{Marginal-based Channel Estimation using Pilots\label{alg:Alg1}}
\begin{enumerate}
\item Initialize $\mathrm{P}\left(h^r_{B}(l)= 1\right)= \lambda_{\mathrm{init}}, \quad l \in \left\{1,2, \cdots, L\right\}, \forall r$.
\item Run \texttt{RS1} at each antenna to estimate its $\bm\lambda$. (Fig. \ref{fig:AlgStep1})
\item\label{en:step1} Each antenna, acting as central antenna, receives marginals from its neighbors. (Fig. \ref{fig:AlgStep2})
\item\label{en:step2} Each antenna computes average marginals $(\bm\lambda')$. (see  (\ref{eq:newmarginals}) and Fig. \ref{fig:AlgStep3})
\item Repeat steps \ref{en:step1}-\ref{en:step2} above, $D$ times. (Fig. \ref{fig:AlgStep4})
\item All antennas re-estimate channels using these marginals as new priors with \texttt{SABMP} algorithm.
\end{enumerate}
\end{algorithm}

\subsection{Algorithm 2: Reduced Communication and Computational Cost -- Integer-based Channel Estimation}
We would like to point out that sharing the marginals vectors among the receiver antennas puts a high communication overhead on the massive-MIMO system. This is because the marginals are floating point numbers and communicating these numbers requires complex signalling. This increases the communication overhead between antennas. However, if just integers are shared, the communication cost could be reduced significantly. We therefore, propose a variant of Algorithm 1 which uses integers for communication among receiver antennas. Since we are not interested in sharing marginals, we do not calculate these and rely on the original \texttt{SABMP} algorithm. Therefore, this algorithm has an additional advantage of low computational complexity as marginals are not calculated.

The algorithm starts by estimating channels at each receiver using \texttt{SABMP} and depends completely on the amplitudes of the estimated MSTs. Following the same reasoning given in the previous section, let $T^{r} = \{  \alpha_1^{r}, \alpha_2^{r}, \cdots, \alpha_{T_{\mathrm{max}}}^{r}\}$ be the set of MSTs of channel $\hv^r$ as detected by \texttt{SABMP} and $\hv^r(T^{r})$ be the corresponding amplitudes. Based on these amplitudes, we define an integer metric for each tap which we call \emph{score} and denote it by $\psi$. For a given $T^{r}$, the highest score is assigned to the channel tap with maximum amplitude in absolute sense. Similarly, the tap (in $T^r$) with minimum amplitude gets the least score. Specifically, since there are $T_{\mathrm{max}}$ MSTs, we assign a score of $T_{\mathrm{max}}$ to the channel tap with maximum amplitude, a score of $T_{\mathrm{max}}-1$ to the second highest tap and so on until a score of $1$ is assigned to the tap with the smallest amplitude among these $T_{\mathrm{max}}$ taps. Therefore, if $|h^r(\alpha_1^r)| > |h^r(\alpha_2^r)| > \cdots > |h^r(\alpha_{T_{\mathrm{max}}}^r)|$ then $\psi(\alpha_1^r)=T_{\mathrm{max}}, \psi(\alpha_2^r)=T_{\mathrm{max}}-1, \cdots, \psi(\alpha_{T_{\mathrm{max}}}^r)=1$ where $\psi(\alpha_i^r)$ is the score of the $i$th detected tap of receiver $r$. All other $L-T_{\mathrm{max}}$ tap locations are assigned a score of zero.

Once each receiver has assigned scores to its detected MSTs, we are ready to invoke the sharing step. Thus, each antenna acting as a central antenna $r_C$ collects these scores from its 4-neighbors and computes the average score for each tap given by
\begin{align}\label{eq:newscores}
\psi(\alpha_i^{r_C}) &= \begin{cases}
    \ceil[\Bigg]{\sum\limits_{j \in \Nc^+} \psi(\alpha_i^j) \Big/ |\Nc^+|}, & \text{if $\alpha_i \in \mathop\bigcup\limits_{j \in \Nc^+}T^j$}\\
    0, & \text{otherwise}
  \end{cases}
\end{align}
where $i\in \{1,2,\cdots,L\}$. The sharing and averaging process is repeated $D$ times and all the related discussion in marginal-based algorithm applies to this algorithm as well. The  averaging step of (\ref{eq:newscores}) is similar to the averaging step of Algorithm 1 given in (\ref{eq:newmarginals}) except that we round up the averaging result to the nearest largest integer. This ensures that the resulting score is always an integer. However, note that the rounding operation is not required in the last step as no sharing takes place after that. Therefore, to avoid unnecessary computation and the resulting information loss, the round up operation is not performed on the average scores. At the end of the $D$ sharing and averaging steps each node computes a belief metric given by
\begin{align}\label{eq:newbeliefs}
b(\alpha_i^r) &= \psi(\alpha_i^r)/T_{\mathrm{max}},
\end{align}
where $b(\alpha_i^r)$ is the estimated belief that the $i$th tap of receiver $r$ is active. Each node uses the beliefs as the Bernoulli priors to re-estimate the channels using \texttt{RS1}. We call this algorithm the \emph{Integer-based Algorithm}. The steps followed by this algorithm are summarized in Algorithm \ref{alg:Alg2}. This algorithm has the following advantages over the marginal-based algorithm:
\begin{enumerate}
\item Reduced communication cost since it totally avoids communicating floating point numbers and,
\item Lower computational complexity since it does not compute marginal probabilities.
\end{enumerate}

\begin{algorithm}
\caption{Integer-based Channel Estimation using Pilots \label{alg:Alg2}}
\begin{enumerate}
\item Run \texttt{SABMP} at each  antenna
\item\label{enIB:step1} Each antenna receives scores from its neighbors
\item\label{enIB:step2} Each antenna computes average scores (\ref{eq:newscores})
\item Repeat steps \ref{enIB:step1}-\ref{enIB:step2} above, $D$ times
\item Each antenna computes a belief vector (\ref{eq:newbeliefs})
\item All antennas re-estimate channels using these belief vectors in place of Bernoulli priors with \texttt{SABMP} algorithm
\end{enumerate}
\end{algorithm}

We now move on to suggest another level of refinement for the marginal probability/scores vectors by selecting reliable data carriers to perform channel estimation.


\section{Data-aided Channel Estimation}\label{sec:data-aided} 

\begin{figure}[htbp]
        \centering
        \begin{tikzpicture}[scale=0.75]
\node (a) at (0,0) {\Large $\pmb\times$};
\node at (0,-5) {\Large $\pmb\times$};
\node at (-5,0) {\Large $\pmb\times$};
\node at (-5,-5) {\Large $\pmb\times$};

\node at (-5+0.5,-5+0.5) {\large $\Xc(c)$};
\node at (-5+5.5,-5+0.5) {\large $\Xc(b)$};
\node at (-5+0.5,-5+5.5) {\large $\Xc(a)$};
\node at (-5+5.5,-5+5.5) {\large $\Xc$};

\draw [black,dashed,thick] (0,0) circle [radius=1.7];

\filldraw (-0.8,-1.5) circle (3pt);
\filldraw (-1.7,0) circle (3pt);

\draw [thick] (-5,0)--(-1.7,0);
\draw [dashed,thick] (-1.7,0)--(0,0);
\draw [thick] (0,-5)--(-0.8,-1.5);
\draw [dashed,thick] (0,0)--(-0.8,-1.5);

\node at (-1.7-0.5,0+0.5) {\large $\widehat\Xc(1) $ \large{ }};
\node at (-0.8-0.5,-1.5-0.5) {\large $\widehat\Xc(2)$};
\end{tikzpicture}
        \caption{Geometrical representation of the reliability measurement. $\widehat{\Xc}(2)$ is more reliable than $\widehat{\Xc}(1)$ as it is less probable to be confused with other constellation points.}\label{fig:GeomRel}
\end{figure}
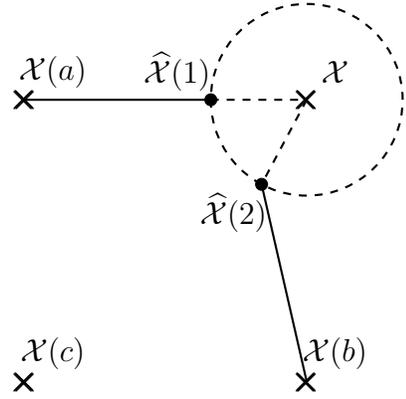

By virtue of the channel sparsity property we can perform channel estimation using a small number of pilots $K$ compared to the channel length $L$ as discussed in the last two sections. We can enhance the channel estimate by increasing the number of pilots. Alternatively, we take a data-aided approach as it is more spectrally efficient. Here, the pilot-based channel estimate is used for data detection which along with the pilots is used to enhance the channel estimate further. Note however that we do not need to use all the detected data for channel estimation thanks to the channel sparsity; a few additional observations would enhance the channel estimate significantly. Therefore, we can be selective and use only the samples which are reliable. So each antenna could independently determine which carrier is reliable by assigning a reliability measure $\mathfrak{R}(i), \; i\in \{ 1, \cdots, N\} \backslash \Pc$ to each of the $N-|\Pc|$ data carriers. That said, we recognize that there are two sources of error in data detection that play important role in determining the reliable data carriers, namely,
\begin{inparaenum}[\itshape a\upshape)] \item noise, and \item error in channel estimation\end{inparaenum}. Their combined distortion effect could be expressed by substituting the estimated channel $\widehat{\hv}_{\rm AMMSE}$ from (\ref{eq:channelerror}) into the system model (\ref{eq:sigmodel}) as follows
\begin{align}
\Ybc &= \Am(\hv+\widetilde{\hv})+\Wbc\nonumber= \Am\hv+\Zbc,
\end{align}
where, $\Zbc = \Am\widetilde{\hv}+\Wbc$ is the combined distortion which is assumed to be Gaussian with zero mean and covariance $\Rm_\Zc$, where $\Rm_\Zc$ is represented in terms of the error covariance $\Rm_{\widetilde{h}}$, calculated in (\ref{eq:error_covariance_matrix}), as
\begin{align}
\Rm_\Zc &= \mathbb{E}[\Zbc \Zbc^\herm]= \mathbb{E}[(\Am\widetilde{\hv} + \Wbc)(\Am\widetilde{\hv} + \Wbc)^\herm]\nonumber\\
&= \mathbb{E}[\Am\widetilde{\hv}\widetilde{\hv}^\herm\Am^\herm + \Wbc\Wbc^\herm]= \Am \Rm_{\widetilde{h}} \Am^\herm + \sigma_w^2\Id.
\end{align}
Here we have assumed that noise $\Wbc$ and error $\widetilde{\hv}$ are uncorrelated. Note that $\Zbc$ includes the effect of both the channel estimation error and the noise and plays the central role in the calculation of reliability measure. Specifically, we use the reliability criterion proposed in \cite{6292976} which takes into consideration the fact that for some carrier $i$, the distortion $\Zc(i)$ might be strong enough to take the estimated data symbol $\widehat{\Xc}(i)$ out of its correct decision region, while for some other carriers the distortion is not strong enough and the data is decoded correctly. All those data carriers $i$ which satisfy this condition $\langle \widehat{\Xc}(i) \rangle = \Xc(i)$, where $\langle \cdot \rangle$ represents rounding to the nearest constellation point, are termed \emph{reliable} carriers and the following metric is used to compute the reliability of carrier $i$
\begin{align}\label{eq:reliability}
\mathfrak{R}(i) & = \frac{p(\Zc(i) = \Xc(i) - \langle \widehat{\Xc}(i) \rangle)}{\sum_{k=0,\Ac_k \ne \langle \widehat{\Xc}(i) \rangle}^{Q-1}p(\Zc(i) = \Xc(i) - \Ac_k)}
\end{align}
where $p(\cdot)$ represents the pdf of $\Zbc$. The numerator in (\ref{eq:reliability}) is the probability that $\Zc(i)$ does not take $\Xc(i)$ beyond its correct decision region and the denominator sums the probabilities of all possible incorrect decisions that $\Zc(i)$ can cause (i.e., $\Zc(i)$ takes $\langle\widehat\Xc(i)\rangle$ to a QAM constellation point $\Ac_k$ different from $\Xc(i)$). The idea of reliability calculation is shown graphically in Fig. \ref{fig:GeomRel}. In this figure, although both $\widehat{\Xc}(1)$ and $\widehat{\Xc}(2)$ are equiprobable to be decoded as $\Xc$ (numerator of (\ref{eq:reliability}) will have same value), $\widehat{\Xc}(2)$ is less likely to be decoded as any other constellation point (denominator of (\ref{eq:reliability}) for $\widehat{\Xc}(2)$ will be smaller) and thus more reliable. Thus, it is obvious that the higher the value of $\mathfrak{R}$ the higher the probability of staying in correct decision region and hence the higher the reliability of the carrier. Note that our reliability calculations of (\ref{eq:reliability}) require the error covariance which is already available at the output of the \texttt{RS1} algorithm as mentioned in Sec. \ref{sec:covar}. Here we would like to point out that the general approach of using reliable data carriers for enhanced channel estimation is not new and techniques employing reliable carriers exist \cite{5419090, 4359544, 1275673}. Specifically, the reliability measure $\mathfrak{R}$ in (\ref{eq:reliability}) is similar to log-likelihood ratios (LLRs) commonly used in joint channel estimation and data detection methods similar to turbo-equalizers (for example, see \cite{1267050} and the references therein).

\begin{algorithm}
\caption{Channel Estimation using Pilots + Reliable Carriers\label{alg:Alg3}}
\begin{enumerate}
\item Run \texttt{Algorithm 1} or \texttt{2} to get CIR estimates
\item Each antenna $r$ uses its estimated channel to find top $U$ reliable carriers $\Rc^r$ and sends $\Rc^r$ to its central antenna
\item Each antenna finds the intersection of received reliable carriers $\Rc = \bigcap_{r\in \Nc^+} \Rc^r$ and sends it back to its neighbors
\item Each antenna sends back data corresponding to $\Rc$ to its central  antenna
\item Each antenna further refines the reliable carriers by selecting only those with same data. Call this list $\Rc^\star$.
\item Each antenna uses the carriers $\Rc^\star$ and the pilots to perform \texttt{SABMP} recovery
\end{enumerate}
\end{algorithm}
Using (\ref{eq:reliability}) each antenna determines the reliability of all data carriers and then select the $U$ carriers with highest reliability values. Let $\Rc^r$ denote the index set of these $U$ reliable carriers for antenna $r$. One possible approach could be that each receiver uses these reliable carriers to enhance the CIR estimate by using Algorithm 1 or 2. However, the antennas can collaborate to enhance the reliability even further. First, each antenna $r_C$ acting as a central antenna collects the indices of the reliable carriers from its 4-neighbors and returns the indices of the reliable carriers common to all antennas, i.e., $\Rc = \bigcap_{r\in \Nc^+} \Rc^r$.
The central antenna can go one step further and ask its neighbors to share their equalized data on the common carriers. The central antenna in turn prunes the set $\Rc$ further and only retains those carriers $\Rc^\star$ on which there is agreement among the neighbors on the value of the transmitted data. The central antenna can now use the enlarged set of pilots plus reliable carriers $\Pc \cup \Rc^\star$ to revisit the channel estimation problem starting from the system of equations
\begin{align}\label{eq:reliabilityprobmodel}
\Ybc^r(\Pc \cup {\Rc^\star}) = \Am(\Pc \cup \Rc^\star) \hv^r + \Wbc^r(\Pc \cup \Rc^\star)
\end{align}
and estimate channel $\hv^r$. The resulting algorithm is presented in Algorithm \ref{alg:Alg3}.

At this stage we would like to point out that implementation of all three algorithms is independent of the antenna array configuration. This is due to the fact that each antenna only deals with its direct neighbors. Therefore, as far as the antennas have the knowledge of their neighbors these algorithms can be implemented on any one-, two-, or three- dimensional arrays with arbitrary topology. Furthermore, the development of algorithms assumed single-antenna UE's; however, the techniques could be easily extended to multiple-antenna UE's such as the LTE UE's which are often equipped with two or four closely located antennas. Since the antennas are closely located, their channels will exhibit the approximately common support property. Therefore, the algorithms explained above could be used to exploit correlation among the channel tap locations to further improve the channel estimates. The only difference is that the effective number of collaborating antennas in each tier will scale with the number of antennas on UE. For example, if there are two transmit antennas on a UE, the number of collaborating antennas in each tier will double.


\section{Simulation Results}\label{sec:results}

\subsection{System Setup}

In this section we will present extensive simulation results to demonstrate the performance of our proposed channel estimation algorithms. Specifically, we consider a MIMO-OFDM system with the simulation parameters given in Table \ref{tab:simparams}.
\begin{table}[htbp]
\centering
{\small
\begin{tabular}{| l | c |}
\hline\hline
\textbf{Parameters} & \textbf{Value}\\
\hline 
Uniform Rectangular Array $(M \times G)$ & $20 \times 20$\\
\hline 
Number of carriers $(N)$& $512$\\
\hline 
Number of pilots $(K)$ & $8, 16$\\
\hline 
QAM modulation order $(Q)$ & $4,16$\\
\hline 
Channel length $(L)$& $32, 64$\\
\hline 
Channel sparsity $(n)$& $\approx 3, 5, 7$\\
\hline 
Collaboration parameter $(D)$& $3$\\
\hline 
\end{tabular}}
\caption{System Parameters for simulation}
\label{tab:simparams}
\end{table}
For simulations, sparse Rayleigh channels are generated where the channel statistics are assumed to be unknown at the receivers. For space-invariant arrays the active tap locations remain fixed across the array. However, for the space-variant case the active tap locations vary slowly across the array. Specifically, we use the IlmProp channel modeling tool \cite{Del_Galdo_ARS_03, IlmProp} for channel generation. It is important to note that there is a general lack of channel models for massive MIMO scenarios and currently IlmProp seems to be one of the best options available to the research community for channel generation. Please refer to Appendix \ref{apd:channelmodel} for relevant discussion. The channels are generated We generate the channels by placing point-like scatterers and the transmitter randomly in the environment and make sure that the line-of-sight is obstructed. Moreover, the number of scatterers is set according to the desired sparsity i.e., $n$. Since the resulting CIR contains many small non-zero components along with the dominant ones, we discard the small ones and keep just the top $n$ components. Further, the center frequency and signal bandwidth are chosen to be $2.6$ GHz and $20$ MHz respectively as specified in the 3GPP-LTE standard. Moreover, to generate the SIA and SVA behavior the distance between antennas was adjusted accordingly.

\subsection{Methods for Performance Comparisons}

The channel vectors $\hv^r$ are estimated using
\begin{inparaenum}[\itshape a\upshape)]
\item least-squares method with known true MST locations (oracle-LS),
\item block-sparse recovery method (BR),
\item proposed Marginal-based channel estimation using pilots (MB-P),
\item proposed Integer-based channel estimation using pilots (IB-P), and
\item proposed Marginal- or Integer-based channel estimation using pilots and reliable carriers (MB-R / IB-R),
\end{inparaenum}

The first two methods are used to benchmark the performance of our algorithm. Oracle-LS knows the channel support at each antenna and hence the only burden is tap estimation using the available pilots. The block sparse recovery method (BR) works in the space-invariant case and uses the fact that the channel support is the same across the array. It casts the problem as several block sparse problems where each receiver collects all observations from its neighbors to estimate the channels. We use the block sparse Bayesian learning algorithm (BSBL) proposed in \cite{BSBL} for block sparse vector estimation as it has been shown to be superior to other methods.

\subsection{Evaluation Criteria}

To evaluate channel estimation performance we use:
\begin{enumerate}
\item Normalized mean-squared error (NMSE) between true and estimated channel vectors.
\begin{equation}
\text{NMSE} = 10 \log_{10} \left( \frac{1}{\Theta} \sum_{\theta=1}^\Theta \frac{\left\| \widehat{\hv}_\theta - \hv_\theta \right\|^2}{\left\| \hv_\theta \right\|^2} \right),
\end{equation}
where $\Theta$ is the number of trials. $\hv_\theta$ and $\widehat{\hv}_\theta$ are the original and estimated CIR at the $\theta$th iteration respectively.
\item Bit-error-rate (BER) between the transmitted data and the recovered data at receivers using the estimated channels. We use zero-forcing equalization to recover the data passed through the channels.

In all of the experiments we average the NMSE and BER over $\Theta=100$ trials.
\end{enumerate}


\subsection{Experiments}

\subsubsection{Experiment 1 - How many pilots?}

\begin{figure*}[htbp]
        \centering
        \begin{tikzpicture}
\begin{axis}[%
width=0.35\textwidth,
height=0.175\textheight,
scale only axis,
xmin=1,
xmax=42,
xlabel={Number of Pilots $(K)$},
xmajorgrids,
ymin=0,
ymax=1,
ylabel={Success Rate},
ymajorgrids,
title style={align=center},
title={\underline{SIA}\\\\ (a)},
]

\addplot [color=red,solid,line width=1.0pt,mark size=3.0pt,mark=|,mark options={solid}]
  table[row sep=crcr]{Exp1ResultsSV0-4.tsv};

\addplot [color=blue,solid,line width=1.0pt,mark size=3.0pt,mark=o,mark options={solid}]
  table[row sep=crcr]{Exp1ResultsSV0-2.tsv};

\addplot [color=brown,solid,line width=1.0pt,mark size=3.0pt,mark=star,mark options={solid}]
  table[row sep=crcr]{Exp1ResultsSV0-3.tsv};

\addplot [color=black,solid,line width=1.0pt,mark size=3.0pt,mark=diamond,mark options={solid}]
  table[row sep=crcr]{Exp1ResultsSV0-1.tsv};

\end{axis}
\end{tikzpicture}%
\begin{tikzpicture}
\begin{axis}[%
width=0.35\textwidth,
height=0.175\textheight,
scale only axis,
xmin=1,
xmax=42,
xlabel={Number of Pilots $(K)$},
xmajorgrids,
ymin=0,
ymax=1,
ylabel={Success Rate},
ymajorgrids,
yticklabel pos=right,
ylabel near ticks,
title style={align=center},
title={\underline{SVA}\\\\ (b)},
legend cell align=left,
legend to name=named,
legend columns=-1,
]

\addplot [color=red,solid,line width=1.0pt,mark size=3.0pt,mark=|,mark options={solid}]
  table[row sep=crcr]{Exp1ResultsSV1-4.tsv};
\addlegendentry{MB-R};

\addplot [color=blue,solid,line width=1.0pt,mark size=3.0pt,mark=o,mark options={solid}]
  table[row sep=crcr]{Exp1ResultsSV1-2.tsv};
\addlegendentry{IB-R};

\addplot [color=brown,solid,line width=1.0pt,mark size=3.0pt,mark=star,mark options={solid}]
  table[row sep=crcr]{Exp1ResultsSV1-3.tsv};
\addlegendentry{MB-P};

\addplot [color=black,solid,line width=1.0pt,mark size=3.0pt,mark=diamond,mark options={solid}]
  table[row sep=crcr]{Exp1ResultsSV1-1.tsv};
\addlegendentry{IB-P};

\end{axis}
\end{tikzpicture}%
\\
\ref{named}
        \caption{Experiment 1: How many pilots are needed to successfully recover the CIR?}\label{fig:sim:Exp1}
\end{figure*}
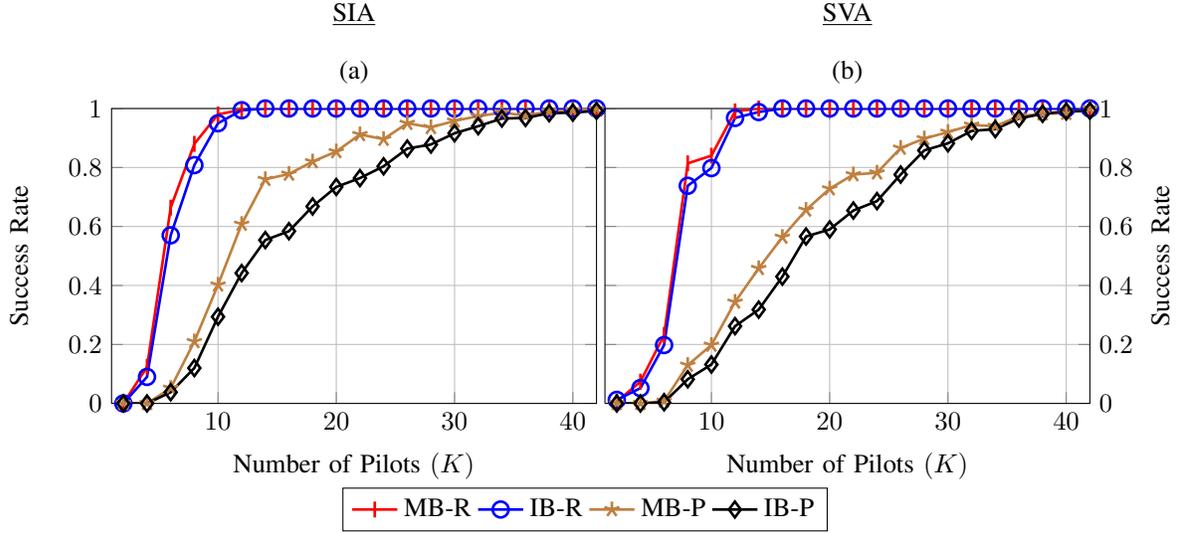

In this experiment, we are interested in finding the required number of pilots for successful recovery of channels of length $L=64$. The graphs in Fig. \ref{fig:sim:Exp1} show the channel recovery success rate vs varying number of pilots for both SIA and SVA.
Note that both pilot-based and data-aided versions of MB and IB algorithms were simulated. Here, success rate is defined as the ratio of the number of successful trials to the number of total trials, where a trial was considered successful when the NMSE was better than $-10$ dB. The SNR was fixed at $10$ dB and the number of pilots $K$ was varied from $2$ to $42$ while $\Theta=100$ trials were conducted for each value of $K$. Channel sparsity was assumed to be $n=3$ and QAM signals of order $Q=4$ were passed through the channels. It is evident from the graphs that for SIA just $6$ pilots are needed by both MB-R and IB-R to achieve a success rate $>  50\%$ and only $12$ pilots to achieve a $100\%$ success rate. This is a small fraction of the channel length $L=64$ (i.e., $9.37\%$ and $18.75\%$ respectively).

\subsubsection{Experiment 2 - Comparison between MB and IB}

\begin{figure*}[htbp]
        \centering
        \begin{tikzpicture}[baseline]
\begin{axis}[%
width=0.25\textwidth,
scale only axis,
xmin=5,
xmax=25,
xmajorgrids,
ymin=-45,
ymax=5,
ylabel={NMSE (dB)},
ymajorgrids,
title style={align=center},
title={\underline{$Q=4$, SIA}\\\\ (a)}
]

\addplot [color=black,solid,line width=1.0pt,mark size=3.0pt,mark=diamond,mark options={solid}]
  table[row sep=crcr]{Exp2ResultsNMSESV0Q4-3.tsv};

\addplot [color=brown,solid,line width=1.0pt,mark size=3.0pt,mark=star,mark options={solid}]
  table[row sep=crcr]{Exp2ResultsNMSESV0Q4-1.tsv};

\addplot [color=blue,solid,line width=1.0pt,mark size=3.0pt,mark=o,mark options={solid}]
  table[row sep=crcr]{Exp2ResultsNMSESV0Q4-4.tsv};

\addplot [color=red,solid,line width=1.0pt,mark size=3.0pt,mark=|,mark options={solid}]
  table[row sep=crcr]{Exp2ResultsNMSESV0Q4-2.tsv};

\end{axis}
\end{tikzpicture}%
\begin{tikzpicture}[baseline]
\begin{axis}[%
width=0.25\textwidth,
scale only axis,
xmin=5,
xmax=25,
xmajorgrids,
ymin=-45,
ymax=5,
ymajorgrids,
yticklabels={,,},
title style={align=center},
title={\underline{$Q=4$, SVA}\\\\ (c)}
]

\addplot [color=black,solid,line width=1.0pt,mark size=3.0pt,mark=diamond,mark options={solid}]
  table[row sep=crcr]{Exp2ResultsNMSESV1Q4-3.tsv};

\addplot [color=brown,solid,line width=1.0pt,mark size=3.0pt,mark=star,mark options={solid}]
  table[row sep=crcr]{Exp2ResultsNMSESV1Q4-1.tsv};

\addplot [color=blue,solid,line width=1.0pt,mark size=3.0pt,mark=o,mark options={solid}]
  table[row sep=crcr]{Exp2ResultsNMSESV1Q4-4.tsv};

\addplot [color=red,solid,line width=1.0pt,mark size=3.0pt,mark=|,mark options={solid}]
  table[row sep=crcr]{Exp2ResultsNMSESV1Q4-2.tsv};

\end{axis}
\end{tikzpicture}%
\begin{tikzpicture}[baseline]
\begin{axis}[%
width=0.25\textwidth,
scale only axis,
xmin=5,
xmax=25,
xmajorgrids,
ymin=-45,
ymax=5,
ylabel={NMSE (dB)},
yticklabel pos=right,
ylabel near ticks,
ymajorgrids,
title style={align=center},
title={\underline{$Q=16$, SIA}\\\\ (e)},
legend cell align=left,
legend to name=named,
legend columns=-1,
]

\addplot [color=black,solid,line width=1.0pt,mark size=3.0pt,mark=diamond,mark options={solid}]
  table[row sep=crcr]{Exp2ResultsNMSESV0Q16-3.tsv};

\addplot [color=brown,solid,line width=1.0pt,mark size=3.0pt,mark=star,mark options={solid}]
  table[row sep=crcr]{Exp2ResultsNMSESV0Q16-1.tsv};

\addplot [color=blue,solid,line width=1.0pt,mark size=3.0pt,mark=o,mark options={solid}]
  table[row sep=crcr]{Exp2ResultsNMSESV0Q16-4.tsv};

\addplot [color=red,solid,line width=1.0pt,mark size=3.0pt,mark=|,mark options={solid}]
  table[row sep=crcr]{Exp2ResultsNMSESV0Q16-2.tsv};

\end{axis}
\end{tikzpicture}%
\vskip 0.25cm
\begin{tikzpicture}[baseline]
\begin{axis}[%
width=0.25\textwidth,
scale only axis,
xmin=5,
xmax=25,
xmajorgrids,
ymode=log,
ymin=0.001,
ymax=1,
ylabel={BER},
ymajorgrids,
title={(b)},
]

\addplot [color=black,solid,line width=1.0pt,mark size=3.0pt,mark=diamond,mark options={solid}]
  table[row sep=crcr]{Exp2ResultsBERSV0Q4-4.tsv};

\addplot [color=brown,solid,line width=1.0pt,mark size=3.0pt,mark=star,mark options={solid}]
  table[row sep=crcr]{Exp2ResultsBERSV0Q4-1.tsv};

\addplot [color=blue,solid,line width=1.0pt,mark size=3.0pt,mark=o,mark options={solid}]
  table[row sep=crcr]{Exp2ResultsBERSV0Q4-5.tsv};

\addplot [color=red,solid,line width=1.0pt,mark size=3.0pt,mark=|,mark options={solid}]
  table[row sep=crcr]{Exp2ResultsBERSV0Q4-2.tsv};

\addplot [color=green,solid,line width=1.0pt,mark size=3.0pt,mark=square,mark options={solid}]
  table[row sep=crcr]{Exp2ResultsBERSV0Q4-3.tsv};

\end{axis}
\end{tikzpicture}%
\begin{tikzpicture}[baseline]
\begin{axis}[%
width=0.25\textwidth,
scale only axis,
xmin=5,
xmax=25,
xlabel={SNR (dB)},
xmajorgrids,
ymode=log,
ymin=0.001,
ymax=1,
ymajorgrids,
yticklabels={,,},
title style={align=center},
title={(d)}
]

\addplot [color=black,solid,line width=1.0pt,mark size=3.0pt,mark=diamond,mark options={solid}]
  table[row sep=crcr]{Exp2ResultsBERSV1Q4-4.tsv};

\addplot [color=brown,solid,line width=1.0pt,mark size=3.0pt,mark=star,mark options={solid}]
  table[row sep=crcr]{Exp2ResultsBERSV1Q4-1.tsv};

\addplot [color=blue,solid,line width=1.0pt,mark size=3.0pt,mark=o,mark options={solid}]
  table[row sep=crcr]{Exp2ResultsBERSV1Q4-5.tsv};

\addplot [color=red,solid,line width=1.0pt,mark size=3.0pt,mark=|,mark options={solid}]
  table[row sep=crcr]{Exp2ResultsBERSV1Q4-2.tsv};

\addplot [color=green,solid,line width=1.0pt,mark size=3.0pt,mark=square,mark options={solid}]
  table[row sep=crcr]{Exp2ResultsBERSV1Q4-3.tsv};

\end{axis}
\end{tikzpicture}%
\begin{tikzpicture}[baseline]
\begin{axis}[%
width=0.25\textwidth,
scale only axis,
xmin=5,
xmax=25,
xmajorgrids,
ymode=log,
ymin=0.001,
ymax=1,
ylabel={BER},
yticklabel pos=right,
ylabel near ticks,
ymajorgrids,
title style={align=center},
title={(f)},
legend cell align=left,
legend to name=named,
legend columns=-1,
]

\addplot [color=black,solid,line width=1.0pt,mark size=3.0pt,mark=diamond,mark options={solid}]
  table[row sep=crcr]{Exp2ResultsBERSV0Q16-4.tsv};
\addlegendentry{IB-P};

\addplot [color=brown,solid,line width=1.0pt,mark size=3.0pt,mark=star,mark options={solid}]
  table[row sep=crcr]{Exp2ResultsBERSV0Q16-1.tsv};
\addlegendentry{MB-P};

\addplot [color=blue,solid,line width=1.0pt,mark size=3.0pt,mark=o,mark options={solid}]
  table[row sep=crcr]{Exp2ResultsBERSV0Q16-5.tsv};
\addlegendentry{IB-R};

\addplot [color=red,solid,line width=1.0pt,mark size=3.0pt,mark=|,mark options={solid}]
  table[row sep=crcr]{Exp2ResultsBERSV0Q16-2.tsv};
\addlegendentry{MB-R};

\addplot [color=green,solid,line width=1.0pt,mark size=3.0pt,mark=square,mark options={solid}]
  table[row sep=crcr]{Exp2ResultsBERSV0Q16-3.tsv};
\addlegendentry{known channel};

\end{axis}
\end{tikzpicture}%
\\
\ref{named}
        \caption{Experiment 2: Performance comparison between marginal-based (MB) and integer-based (IB) algorithms.}\label{fig:sim:Exp2}
\end{figure*}
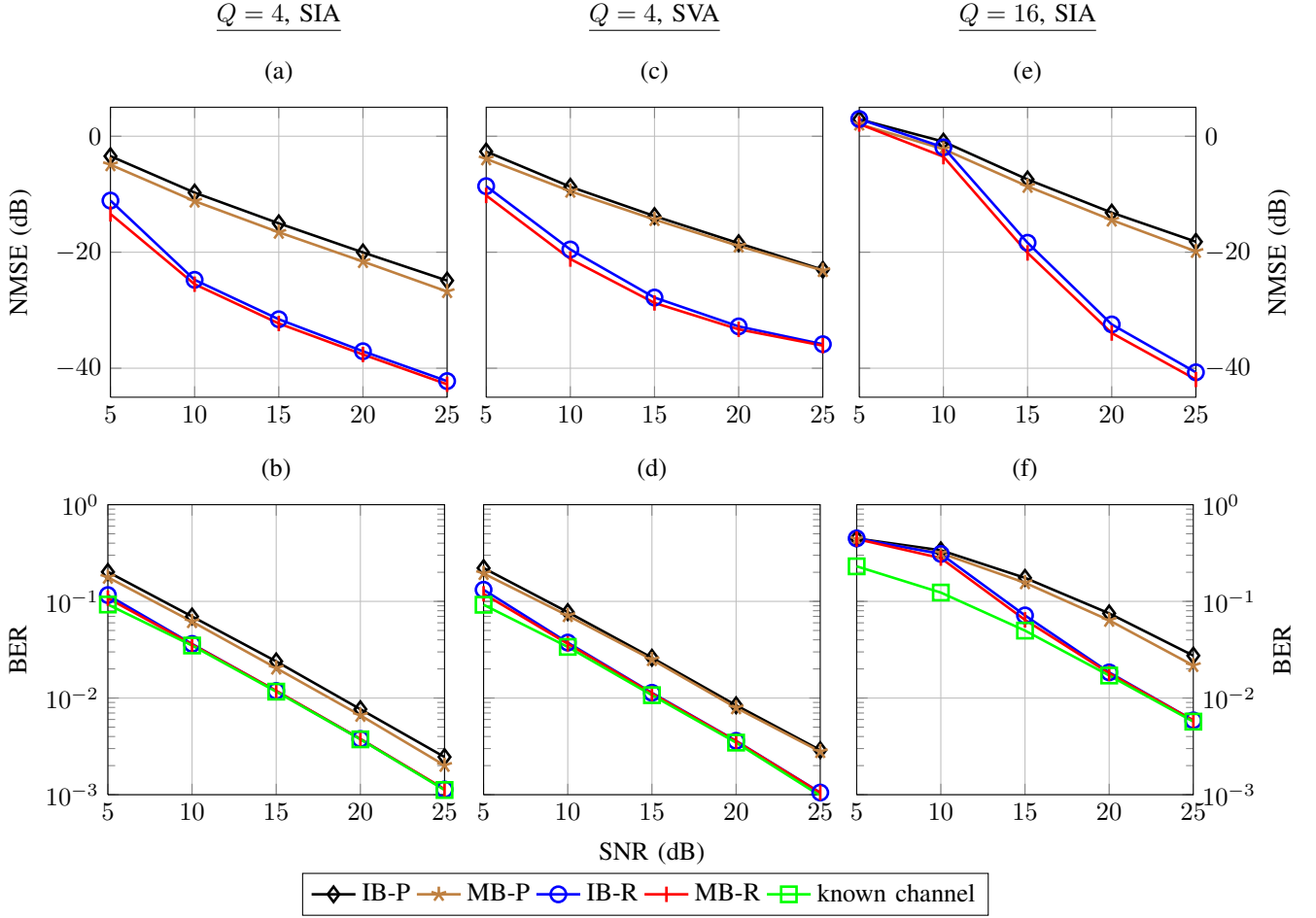

In this experiment, we compare the performance of the proposed MB and IB channel estimation algorithms. Channels of length $L=64$ and sparsity $n=3$ were estimated using $K=16$ pilots. The top row of Fig. \ref{fig:sim:Exp2} shows the NMSE in estimated CIRs while the bottom row shows the BER of the recovered data using these CIR estimates. Moreover, as apparent from the labels, this experiment was run using three different choices of parameters, namely, ($Q=4$, SIA), ($Q=4$, SVA), and ($Q=16$, SIA) respectively. The figure shows that incorporating reliable carriers results in significant performance gains. The figure also shows that the algorithms perform equally well for both SIA and SVA configurations and, hence are robust to how accurate property 2 is.

An important observation is that there is little advantage of MB algorithms over IB algorithms in this setting. However in some scenarios the improvement could be better as it uses more (and accurate) information to estimate the channels. For example, the difference between the performance of MB and IB algorithms is more evident when the number of pilots is further reduced, as could be seen in the success rate curves of Fig. \ref{fig:sim:Exp1}. Also note that the gain of MB algorithms is much more noticeable for their pilot-based versions. However, this advantage is at the expense of a relatively high computational and communication cost as depicted in Table \ref{tab:sim:PBvsIB_runtime} which compares the runtime for the two algorithms.
\begin{table}[htbp]
\centering
{\small
\begin{tabular}{|c|c|}
\hline\hline
\textbf{Algorithm} & \textbf{Run time (sec)}\\
\hline 
MB-P & 0.3897\\
\hline 
IB-P & 0.3095\\
\hline 
\end{tabular}}
\caption{Run time comparison of the MB-P and IB-P.}
\label{tab:sim:PBvsIB_runtime}
\end{table}

\subsubsection{Experiment 3 - Comparison with BR and oracle-LS algorithms}%
In this experiment, we benchmark the performance of the proposed algorithms against BR and oracle-LS. Here we use QAM signals of order $Q=4$ and $Q=16$ and pass them through a channel of length $L=32$ and sparsity $n=3$. We use $K=8$ pilots and confine our attention to the space-invariant case as this is an essential requirement for block sparsity algorithm to work.
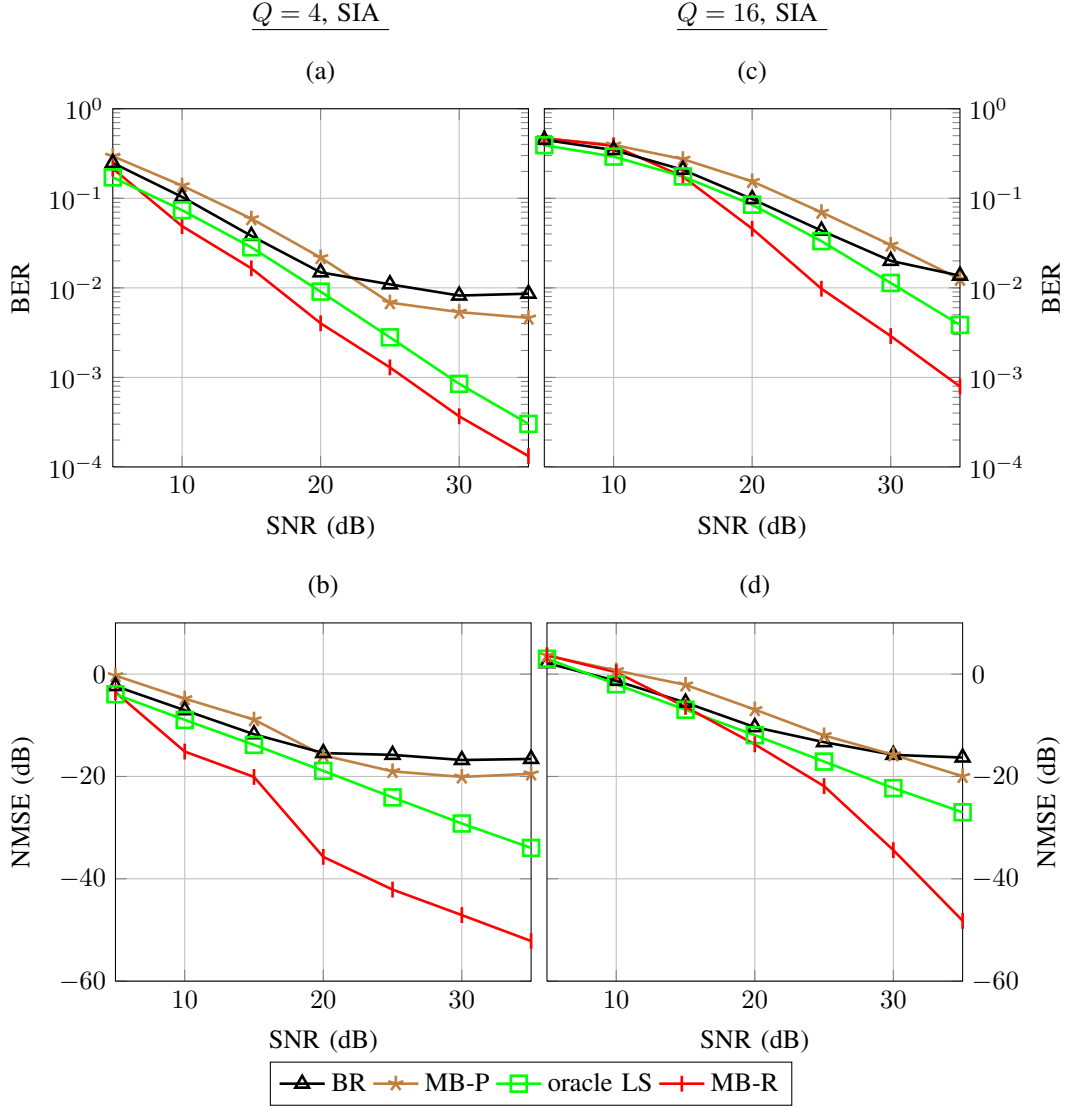
\begin{figure*}[htbp]
        \centering
        \begin{tikzpicture}

\begin{axis}[%
width=0.3\textwidth,
scale only axis,
xmin=5,
xmax=35,
xlabel={SNR (dB)},
xmajorgrids,
ymode=log,
ymin=0.0001,
ymax=1,
yminorticks=true,
ylabel={BER},
ymajorgrids,
title style={align=center},
title={\underline{$Q=4$, SIA }\\\\ (a)},
legend style={draw=black,fill=white,legend cell align=left}
]
\addplot [color=brown,solid,line width=1.0pt,mark size=3.0pt,mark=star,mark options={solid}]
  table[row sep=crcr]{Exp3BERQ4-1.tsv};

\addplot [color=red,solid,line width=1.0pt,mark size=3.0pt,mark=|,mark options={solid}]
  table[row sep=crcr]{Exp3BERQ4-2.tsv};

\addplot [color=black,solid,line width=1.0pt,mark size=3.0pt,mark=triangle,mark options={solid}]
  table[row sep=crcr]{Exp3BERQ4-3.tsv};

\addplot [color=green,solid,line width=1.0pt,mark size=3.0pt,mark=square,mark options={solid}]
  table[row sep=crcr]{Exp3BERQ4-4.tsv};

\end{axis}
\end{tikzpicture}%
\begin{tikzpicture}

\begin{axis}[%
width=0.3\textwidth,
scale only axis,
xmin=5,
xmax=35,
xlabel={SNR (dB)},
xmajorgrids,
ymode=log,
ymin=0.0001,
ymax=1,
yminorticks=true,
ylabel={BER},
ymajorgrids,
yticklabel pos=right,
ylabel near ticks,
title style={align=center},
title={\underline{$Q=16$, SIA }\\\\ (c)},
legend style={draw=black,fill=white,legend cell align=left}
]
\addplot [color=brown,solid,line width=1.0pt,mark size=3.0pt,mark=star,mark options={solid}]
  table[row sep=crcr]{Exp3BERQ16-1.tsv};

\addplot [color=red,solid,line width=1.0pt,mark size=3.0pt,mark=|,mark options={solid}]
  table[row sep=crcr]{Exp3BERQ16-2.tsv};

\addplot [color=black,solid,line width=1.0pt,mark size=3.0pt,mark=triangle,mark options={solid}]
  table[row sep=crcr]{Exp3BERQ16-3.tsv};

\addplot [color=green,solid,line width=1.0pt,mark size=3.0pt,mark=square,mark options={solid}]
  table[row sep=crcr]{Exp3BERQ16-4.tsv};

\end{axis}
\end{tikzpicture}%
\vskip 0.25cm
\begin{tikzpicture}

\begin{axis}[%
width=0.3\textwidth,
scale only axis,
xmin=5,
xmax=35,
xlabel={SNR (dB)},
xmajorgrids,
ymin=-60,
ymax=10,
ylabel={NMSE (dB)},
ymajorgrids,
title={(b)},
legend style={draw=black,fill=white,legend cell align=left}
]
\addplot [color=brown,solid,line width=1.0pt,mark size=3.0pt,mark=star,mark options={solid}]
  table[row sep=crcr]{Exp3NMSEQ4-1.tsv};

\addplot [color=red,solid,line width=1.0pt,mark size=3.0pt,mark=|,mark options={solid}]
  table[row sep=crcr]{Exp3NMSEQ4-2.tsv};

\addplot [color=black,solid,line width=1.0pt,mark size=3.0pt,mark=triangle,mark options={solid}]
  table[row sep=crcr]{Exp3NMSEQ4-3.tsv};

\addplot [color=green,solid,line width=1.0pt,mark size=3.0pt,mark=square,mark options={solid}]
  table[row sep=crcr]{Exp3NMSEQ4-4.tsv};

\end{axis}
\end{tikzpicture}%
\begin{tikzpicture}

\begin{axis}[%
width=0.3\textwidth,
scale only axis,
xmin=5,
xmax=35,
xlabel={SNR (dB)},
xmajorgrids,
ymin=-60,
ymax=10,
ylabel={NMSE (dB)},
ymajorgrids,
yticklabel pos=right,
ylabel near ticks,
title={(d)},
legend to name=named,
legend columns=-1,
]
\addplot [color=black,solid,line width=1.0pt,mark size=3.0pt,mark=triangle,mark options={solid}]
  table[row sep=crcr]{Exp3NMSEQ16-3.tsv};
\addlegendentry{BR};

\addplot [color=brown,solid,line width=1.0pt,mark size=3.0pt,mark=star,mark options={solid}]
  table[row sep=crcr]{Exp3NMSEQ16-1.tsv};
\addlegendentry{MB-P};

\addplot [color=green,solid,line width=1.0pt,mark size=3.0pt,mark=square,mark options={solid}]
  table[row sep=crcr]{Exp3NMSEQ16-4.tsv};
\addlegendentry{oracle LS};

\addplot [color=red,solid,line width=1.0pt,mark size=3.0pt,mark=|,mark options={solid}]
  table[row sep=crcr]{Exp3NMSEQ16-2.tsv};
\addlegendentry{MB-R};

\end{axis}
\end{tikzpicture}%
\\
\ref{named}
        \caption{Experiment 3: Performance comparison between the proposed and the BR and oracle-LS algorithms.}\label{fig:sim:allcomparison}
\end{figure*}

It is obvious from the graphs of Fig. \ref{fig:sim:allcomparison} that the proposed MB-R algorithm has the best performance among all algorithms.
The gain in the performance of the proposed algorithm over others is more prominent for higher values of SNR. Specifically, note that there is a difference of nearly two orders of magnitude in the BER of MB-R and BR when SNR $=35$ dB and $Q=4$.%

\subsubsection{Experiment 4 - Effect of sparsity rate}

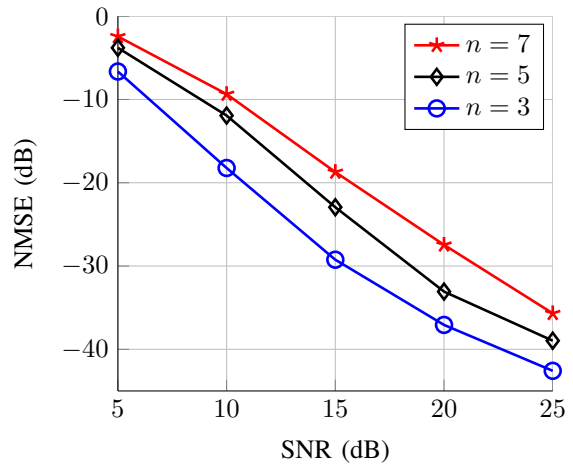
\begin{figure}[htbp]
        \centering
        \begin{tikzpicture}

\begin{axis}[%
width=0.65\columnwidth,
scale only axis,
xmin=5,
xmax=25,
xlabel={SNR (dB)},
xmajorgrids,
ymin=-45,
ymax=0,
ylabel={NMSE (dB)},
ymajorgrids,
axis x line*=bottom,
axis y line*=left,
legend style={draw=black,fill=white,legend cell align=left}
]
\addplot [color=red,solid,line width=1.0pt,mark size=3.0pt,mark=star,mark options={solid}]
  table[row sep=crcr]{Exp4Results-3.tsv};
\addlegendentry{$n=7$};

\addplot [color=black,solid,line width=1.0pt,mark size=3.0pt,mark=diamond,mark options={solid}]
  table[row sep=crcr]{Exp4Results-2.tsv};
\addlegendentry{$n=5$};

\addplot [color=blue,solid,line width=1.0pt,mark size=3.0pt,mark=o,mark options={solid}]
  table[row sep=crcr]{Exp4Results-1.tsv};
\addlegendentry{$n=3$};

\end{axis}
\end{tikzpicture}%
        \caption{Experiment 4: Effect of channel sparsity on its recovery.}\label{fig:sim:EffectOfSparsityRate}
\end{figure}

In this experiment, we study the performance of the proposed algorithms under different sparsity rates. Channels of length $L=64$ were generated having $n=3,5$ and $7$ non-zero taps corresponding to sparsity rate of $4.7\% - 11\%$. In this experiment, QAM signals of order $Q=4$ were passed through the channels and $K=16$ number of pilots were used. Fig. \ref{fig:sim:EffectOfSparsityRate} shows the NMSE performance of the proposed IB-R algorithm. It is evident from the graphs that the performance of the proposed algorithm degrades gracefully as CIR gets denser. A similar performance is achieved for the MB-R algorithm. The degradation in reconstruction accuracy with increased number of non-zeros is a common trend in all sparse recovery algorithms.%

\subsubsection{Experiment 5 - Effect of $D$}\label{sec:Exp5}

\begin{figure}[htbp]
        \centering
        \includegraphics[width=1.15\columnwidth]{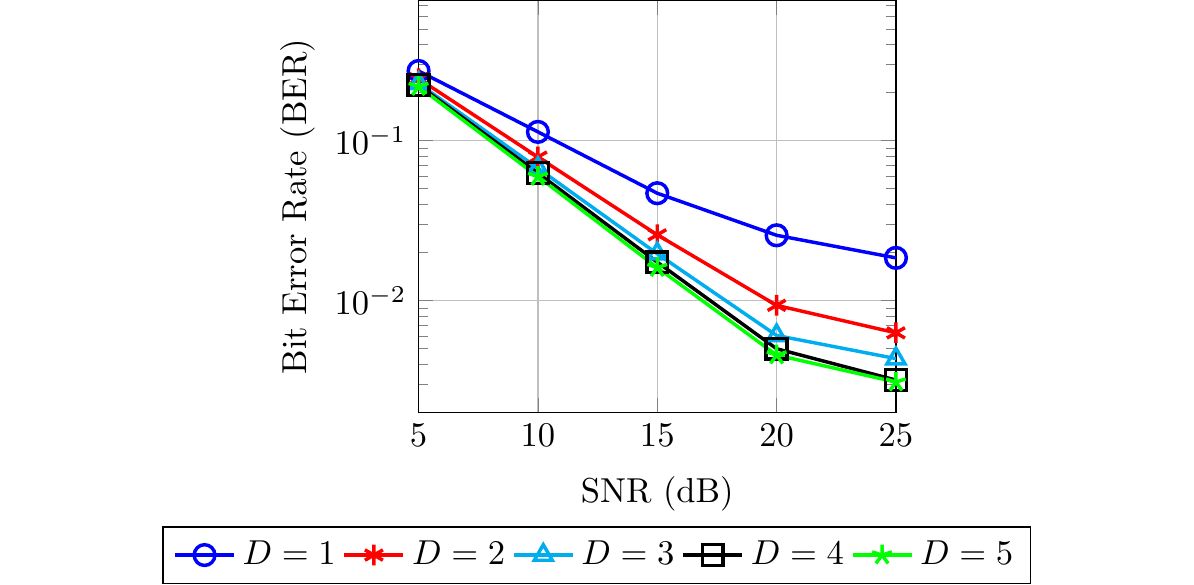}
        \caption{Experiment 5: Information sharing among antennas belonging to different neighbor levels ($D = 1 - 5$) adds to CIR estimation accuracy in the SIA case.}\label{fig:sim:ChoiceOfDSIA}
\end{figure}

\begin{figure}[htbp]
        \centering
        \includegraphics[width=1.15\columnwidth]{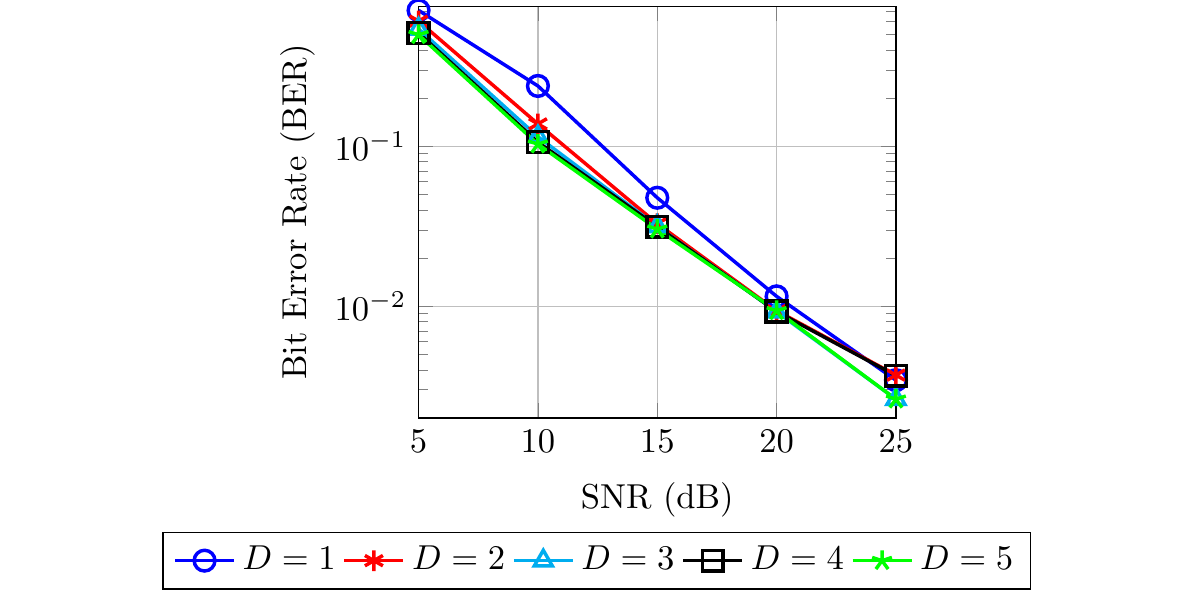}
        \caption{Experiment 5: In the SVA case information sharing does not help in the improvement of CIR estimation accuracy.}\label{fig:sim:ChoiceOfDSVA}
\end{figure}

In this experiment, we study the effect of the number of collaborating antennas on channel estimation for both the SIA and SVA cases. In this experiment we compute the BER in the recovered data for various values of SNR and parameter $D$. Fig. \ref{fig:sim:ChoiceOfDSIA} and \ref{fig:sim:ChoiceOfDSVA} show the BER vs SNR graphs for SIA and SVA respectively. Specifically, we plot for $D=1,2,3,4$ and $5$. Note that when $D=1$, information from only the direct 4 neighbors is taken into consideration and for $D=2$ information belonging to the neighbors of neighbors is also incorporated in the computations.

In the SIA case (Fig. \ref{fig:sim:ChoiceOfDSIA}) QAM signals of order $Q=4$ were passed through channels of length $L=32$ and sparsity $n=3$ and the corresponding CIRs were estimated using IB-P with the help of $K=8$ pilots. Fig. \ref{fig:sim:ChoiceOfDSIA} shows that sharing improves the BER performance. Specifically, as we increase the scope of sharing from the first tier of neighbors ($D=1$) to the fifth tier ($D=5$), we observe a drop in BER. However, the improvement in BER is not significant beyond $D=3$. This behavior is dependent on several factors such as the length of channel ($L$) to be estimated, number of pilots ($K$) and the channel sparsity ($n$). For instance, in this example, if the number of pilots is reduced (i.e., $K<8$) the estimated CIRs will be more erroneous. However, the effect of this error is compensated by using more neighbors to average the marginals/scores (i.e., higher $D$). Therefore, in this reduced number of pilots scenario we might observe significant improvement beyond $D=3$ as well.

In the SVA case (Fig. \ref{fig:sim:ChoiceOfDSVA}) QAM signals of order $Q=4$ were passed through channels of length $L=64$ and sparsity $n=3$ and the corresponding CIRs were estimated using IB-P with the help of $K=16$ pilots. Fig. \ref{fig:sim:ChoiceOfDSVA} shows that we do not gain anything for higher values of $D$. This is due to the space-variant nature of the impulse response that adding more information does not help in the improvement of estimation accuracy. Therefore, setting $D=1$ or $2$ might be sufficient in this scenario.

\section{Conclusion and Future Work}\label{sec:conclusions}
Massive MIMO systems provide substantial performance gains as compared to the traditional MIMO systems. However, these gains come with a huge requirement of estimating a large number of channels. In this paper we have shown that these channels could be estimated in a collaborative manner where the antennas collaborate with their neighboring antennas. Three algorithms based on this collaborative method have been presented. The algorithms show good performance under different scenarios and that too while using a relatively small number of pilots.


\appendices

\section{Channel Models}\label{apd:channelmodel}
Numerous good/accurate models exist for MIMO wireless channels. However, almost all consider the uniform linear array configuration of the antennas \cite{chan_mod_bolcskei, SCM_3GPP}. It should be noted that from a practical point of view, assuming a uniform linear array (ULA) becomes unfeasible for the purpose of modeling a large scale antenna array. Therefore, the use of two- or three-dimensional antenna arrays can be more appropriate in massive MIMO systems. There have been a few attempts in developing channel models for the two- and three-dimensional antenna array configurations. For example, the IlmProp tool available online \cite{IlmProp} allows to generate CIR for 2D antenna array configuration. A similar proposal was put forward in \cite{boon_2D_antennaarray, 3D_chann_proposal}, to extend the spatial channel model (SCM) standard \cite{SCM_3GPP}. Similar models have also been considered in Winner II \cite{WinnerII} and Winner+ \cite{Winner_plus} initiatives. However, these models come with their limitations. For example, in IlmProp the parameters have been estimated using much lower spatial resolution. Indeed, there is a general lack of channel models for the massive MIMO scenarios and there is a  need of more work in this direction.

\bibliographystyle{IEEEtran}
\bibliography{IEEEabrv,MyLib3}

\end{document}